# Key Stakeholders' Value Propositions for Feature Selection in Software-intensive Products: An Industrial Case Study

Pilar Rodríguez, Emilia Mendes, and Burak Turhan, *Member, IEEE*

**Abstract**— Numerous software companies are adopting value-based decision making. However, what does value mean for key stakeholders making decisions? How do different stakeholder groups understand value? Without an explicit understanding of what value means, decisions are subject to ambiguity and vagueness, which are likely to bias them. This case study provides an in-depth analysis of key stakeholders' value propositions when selecting features for a large telecommunications company's software-intensive product. Stakeholders' value propositions were elicited via interviews, which were analyzed using Grounded Theory coding techniques (open and selective coding). Thirty-six value propositions were identified and classified into six dimensions: customer value, market competitiveness, economic value/profitability, cost efficiency, technology & architecture, and company strategy. Our results show that although propositions in the customer value dimension were those mentioned the most, the concept of value for feature selection encompasses a wide range of value propositions. Moreover, stakeholder groups focused on different and complementary value dimensions, calling to the importance of involving all key stakeholders in the decision making process. Although our results are particularly relevant to companies similar to the one described herein, they aim to generate a learning process on value-based feature selection for practitioners and researchers in general.

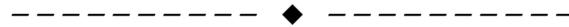

## 1 INTRODUCTION

SOFTWARE Engineering (SE) research has historically stressed the economic aspect of software-related decisions (e.g. [1], [2]), where short-term aspects such as project costs and schedule were predominantly considered (i.e. *'every requirement, use case, object, and defect is treated as equally important'* – from a value perspective [3]). In today's cutthroat product and services industries, software has become the main driver for competitive advantage [4]. As the size and complexity of software-based solutions increase, so does the impact of software development decisions on the overall product offering [4]. The need to balance short and long-term aspects motivated the emergence of a research area called Value-Based SE (VBSE) [3], [5], which stresses thinking value in software-related decisions. VBSE highlights that it is not enough to merely meet schedules, budget and quality objectives; ultimately, software needs to create value [5]. Thus, value represents the inclusion of both short and long-term aspects when making software-related decisions. The recent popularity of agile methods, and its emphasis on delivering value to customers [6], [7], have contributed to move the VBSE agenda forward, enlarging its scope [5]. Numerous companies embrace today the paradigm shift led by the VBSE agenda, placing value at center stage in the SE landscape.

However, despite the unquestionable importance that value, as a concept, has in SE, its detailed understanding has not yet been forthcoming. For example, although value propositions (i.e. criteria/factors used to determine the value of an item) are often referenced as an integral part of SE decision making, they are poorly documented and rarely explicitly utilized, even within development processes that emphasize value (e.g. agile [8]). As a result, decision making is usually carried out based solely upon implicit and tacit value propositions, being subject to ambiguity and vagueness, which are likely to bias decisions given insufficient reasoning about criteria important to create value (e.g. errors of omission) or conflicting views among stakeholders [5].

Existing research (e.g. [9], [10], [11]) supports such limited view of value. A systematic mapping study investigating value aspects in SE and other fields [9] suggests that the research contributions in the SE area are often isolated and with a limited choice of value aspects; for example, focusing only on cost, or on product characteristics such as usability. Using as basis the work in [9] and additional literature from economics, business, and marketing research, Khurum et al. [4] proposed the software value map (SVM), a large classification of 62 value propositions (called components), that aims to be general enough to represent the views of all stakeholders who make any decision related to any software product. However, its general nature makes it ambiguous to apply in specific situations (e.g. feature selection). Moreover, the SVM does not look into how differ-

————————————————

- *Pilar Rodríguez and Emilia Mendes are with the University of Oulu, Finland, E-mail: pilar.rodriguez@oulu.fi, emilia.mendes@oulu.fi*
- *Emilia Mendes is with the Blekinge Institute of Technology, Sweden. E-mail: emilia.mendes@bth.se*
- *Burak Turhan is with Monash University, Australia. E-mail: burak.turhan@monash.edu*

ent companies interpret value and which value propositions are truly important in practice [10]. The few empirical studies on the topic (e.g. [10], [12], [13]), which are further developed in Sections 2 and 5, provide preliminary insights into the concept of value. However, their findings are limited due to shortcomings with the research method used (e.g. survey studies) or due to their narrow detailing of value (e.g. not including value propositions arranged by key stakeholder groups).

The aforementioned shortcomings are addressed in this paper, where we present a case study [14] conducted with a large telecommunication company. The study focuses on the concept of value in the context of feature selection for release planning. Features represent needs that are gathered via meetings with customers or other stakeholders [15], which, once selected, are refined during a requirements elicitation process [16], [17]. Concretely, we analyzed the value propositions that key stakeholders use when deciding upon software features for release planning in our company case in order to answer the following research questions (RQs):

*RQ1. What are the value propositions (value factors) employed by key stakeholders when making decisions relating to feature selection for release planning?*

*RQ2. What are the overlapping and distinct value propositions (value factors) between the different stakeholder groups involved in value-based feature selection?*

Data was collected through interviews with all ten key stakeholders involved in feature selection for the product in the focus of this study, including strategic product managers, product owners, project managers and technical leaders. By using grounded theory (GT)'s coding techniques [18], [19], we translated stakeholders' tacit knowledge related to value propositions into an explicit and tangible value representation. Stakeholders' value propositions were next reconciled and differences between stakeholder groups were analyzed. The outcome of this study is a consolidated view of value through a detailed analysis of value propositions deemed important when selecting features in our company case.

The main contributions of the study are as follows:
- The results contribute to the area of VBSE and, specifically, value-based feature selection by providing a detailed set of empirical data on value propositions that decision-makers use when selecting features for release planning. The focus on feature selection means that the elicited propositions are more specific, with clear use cases, as compared to related work.
- Compared to previous research, we provide a more detailed understanding of value propositions empirically identified, i.e., we provide proper definitions and detailed descriptions. In addition, we also identify value propositions that have been overlooked by existing empirical research but are nevertheless important.
- The per-stakeholder analysis of value is a new contribution, in contrast to previous work. The findings show that stakeholder groups provide different but complementary value propositions. Although there were not strong conflicting views between stakeholder groups, value propositions did not completely overlap, calling for the importance of involving all key stakeholders in the decision-making process.
- The detailed comparison to related studies, presented in Section 5, is also, as far as we know, a unique contribution of this study. Results show agreement with previously found value propositions, but also reveal several new propositions. Value propositions seem to be diverse and many of them context dependent.
- With regard to a company case's perspective, the consolidated view of value can be used to develop a common understanding of value, taking away subjectivity and misunderstandings, and as a decision support vehicle as it includes all value propositions accounted for when selecting features.

Researchers can benefit from the study not only by deepening their understanding of the role of value propositions in SE, but also using it as a framework to enable further research and as basis from which more general theory could emerge. Moreover, the value propositions identified in this study can be used as an input for existing release planning methods, such as the VALUE framework (see Section 2.2). From a practitioners' perspective, the study provides an exemplar that can be useful in similar software projects and company contexts. Further, the detailed knowledge provided on value propositions and stakeholder analysis may be useful for practitioners in general who want to learn about value-based feature selection.

The rest of the paper is structured as follows: Section 2 presents background and related work. The case study design is described in Section 3. Section 4 presents the results, which are discussed in light of earlier studies on the topic in Section 5, including implications for research and practice. Threats to validity are also discussed in Section 5. Finally, Section 6 presents conclusions.

## 2 BACKGROUND AND RELATED WORK

This section presents background literature deemed useful for the reader to understand our study and the research gaps. The literature is further developed in Section 5, when our results are discussed in light of related work.

### 2.1 Background

Over the past fifty years value has been described in a variety of ways. Miles [20] sees value as an equation where *Customer* value is calculated as a function of performance (how fast a product is produced) by cost; therefore, value can be increased by either improving performance or reducing costs. His proposal led to several value models that enabled different functions to be compared (e.g. [21], [22]). For example, Shillito and De Marle [22] defined value as the product of the need for an object, times its ability to satisfy such need divided by the object's cost. A detailed historical account on value in different domains is provided in [23] and [24].

The economics of software has been a relevant topic of study in SE for years (e.g. [1], [25], [26]). A main research focus has been upon effort and cost measurement and estimation, rather than to consider value [5], which has led to a large body of knowledge on cost estimation techniques (e.g. [27], [28]). There is also a wide body of knowledge on requirements/features prioritization, including methods and models for release planning (e.g. [29], [30], [31]), requirements prioritization techniques [32] and studies on managing dependencies between requirements/features (e.g. [33], [34]). Although only a few prioritization techniques have explicitly considered value [32], this body of knowledge has not focused only on cost, effort, or schedule but has often included an inherent dimension of importance, based on complex multi-criteria, which resembles value [32]. Thus, the work on requirements prioritization has not been value-neutral. Still, its main focus has not been on understanding value or providing empirical evidence on value, leaving the notion of value vague or implicit [10], [13]. Differently, our study focuses on understanding explicitly the value propositions that stakeholders employ when selecting features for a release. Thus, our work complements past work on requirements prioritization as the value propositions elicited in this study could feed into some existing prioritization techniques.

Boehm's paper [3] can be considered the starting point for VBSE. In this call to arms, Boehm put forward that much of the SE practice and research at that time was carried out in a value neutral setting, where '*"Earned value" systems tracked project cost and schedule, not stakeholder or business value*' and '*a "separation of concerns" was practiced, in which the responsibility of software engineers was confined to turning software requirements into verified code*'. Boehm [35] warned that, "*most studies of the critical success factors distinguishing successful from failed software projects find that the primary critical success factors lie in the value domain*". He also identified nine important areas of the software development process that are relevant from a value perspective, including value-based requirements engineering, which is the focus of our study, as feature selection is part of that area. Further, Boehm and Jain [36] developed the 4+1 theory of VBSE, which stresses identifying which value propositions are important for success-critical stakeholders, so to achieve and maintain a win-win state.

Boehm made it clear that value comprises different propositions (factors), which are defined in terms of project requirements and company goals/objectives. However, an explicit list of value propositions is not provided. Therefore, one of the challenges in VBSE is to ascertain these value propositions from key stakeholders who make value-based decisions, so to improve our understanding relating to value-based decision making in SE and to provide companies with a consolidated view of value that can be used to develop a common understanding, taking away subjectivity and misunderstandings.

Boehm's seminal paper encouraged others in the SE community to focus on value. Over time, the scope of VBSE research expanded to integrate value-oriented perspectives other than economic and monetary aspects [5]. Around the same time, the agile community also started to stress the importance of delivering valuable software to customers [37]. Agile and Lean software development base their views of value mainly on a customer perspective (i.e., features are selected based on what is expected to provide higher customer value) [38], [39]. More recent software development processes, such as continuous deployment [40], also emphasize a *value driven real-time business* paradigm. These are examples supporting the view that value became center stage in the SE landscape.

## 2.2 Related Work

Existing empirical research on value propositions is relatively scarce and lacks detail to understand the meaning of these propositions. Besides the secondary studies presented in Section 1 ([4], [9], [41]), there have been some empirical studies investigating stakeholders value propositions in relation to product management (studies close to the scope of this research) (e.g. [10], [12], [13], [42]-[46]). However, these studies, which are further discussed in Section 5, differ from our approach in three distinct ways:

i) Most previous studies employed a top-down approach by means of research methods such as surveys based on semi-structured questionnaires, and using checklists as data collection instruments. As a consequence, value propositions are predefined by researchers, instead of emerging in collaboration with key stakeholders. Such approaches can lead to results that miss important criteria due to issues such as anchoring effect [47]. Moreover, they lack details for the identified criteria as findings are mainly based on closed-ended questions and can also suffer from a range of problems, such as nonresponse bias [48].

ii) Previous studies did not consider all key stakeholders involved in the studied decision process. For example, business stakeholders seldom participated in previous studies, which led to value propositions from such critical stakeholder group being absent or under-represented.

iii) A few studies base their results solely on notes taken during interviews and analyzed using thematic analysis. Such approach does not guarantee that all important knowledge has been elicited, as it depends upon notes taken, rather than on complete transcripts from recorded interviews [49].

Our work, in turn, has used the text from transcribed interviews with all key stakeholders involved in feature selection in our case, in addition to GT coding techniques [18] to identify and analyze value propositions. Such method enabled key stakeholders to speak freely (vs. group interviews that may hinder identifying conflicting views among stakeholders) and for the researchers to focus on the knowledge being provided. GT coding techniques allowed going deeply into the details of each identified value proposition [18]. This is the first time such approach is employed to elicit value propositions from key stakeholders.

Related to our own previous work on the topic, this study is part of a larger research effort, the VALUE framework, which aims to improve decision-making in software/software-intensive development and management. Its vision was introduced in [50] and detailed in [51]. It includes a mixed-methods approach where key stakeholders' tacit knowledge regarding propositions used during decision-making are elicited and used as input to a Web-based tool (Value tool). Such tool is then used by those key stakeholders to support their decision-making process and build a value estimation Bayesian Network model [51]. Our previous work details the use of the Value tool for decision-making [52], [53] and the building of value estimation Bayesian Network models [54]. The research presented herein is the first to detail the explicitation of tacit knowledge from key stakeholders.

TABLE 1
COMPANY CASE - CONTEXT INFORMATION (ADAPTED FROM PETERSEN AND WOHLIN [58])

| Context facet | Context element | Description |
|---|---|---|
| Product | Quality | Regulated domain (specific telecommunication regulations and standards) |
| | Size | Commercial product, large-scale system |
| | System type | Embedded system |
| | Customization | Tailored product to different market segments |
| Process | Overall software development process | Scrum-like development process with continuous and incremental development of features. Features are written in natural language and managed using product backlogs |
| Practices, tools and techniques (for feature selection) | Feature selection practices and techniques | - Value-based feature selection<br>- Iterative feature-selection process with periodic decision making meetings: weekly distributed meetings and around 2-3 face-to-face meetings per release<br>- Features managed using two product backlogs: Company Product Backlog (CPB) – Company's headquarters, and Development Product Backlog (DPB) – Development unit |
| | Decision making tool support | Commercial tool used to manage the product backlog at company level (CPB). MS Excel sheets used to keep the product backlog in the development unit (DPB) |
| People (feature selection team) | Roles involved in feature selection | - Three strategic product managers located at the company headquarters<br>- Product owner, project managers and technical stakeholders at the development unit |
| | Their experience | Wide experienced stakeholders, from 27 to 10 years of experience in software development. The strategic product managers have been working at the company for roughly 20 years |
| Organization | Size | Large organization |
| | Organizational model | Several locations distributed worldwide. Flat organization based on Agile principles |
| | Organizational unit (part of the organization involved in the study) | - Company's headquarters (strategic product management)<br>- Development unit (in charge for developing the product. The unit is composed by around 300 people and has high expertise on the technologies used for developing the product) |
| | Distribution | System locally developed in the development unit but decision making team partially distributed in two locations (company headquarters and development unit, which is in another country) |
| Market | Number of customers | Market-driven software development |
| | Setting | Business-to-business (B2B) |
| | Constraints | Market characterized by uncertainty and short time-to-market |

## 3 CASE STUDY DESIGN

This research is classified as applied research since its main contribution is the practical knowledge generated related to value propositions for value-based feature selection [55]. We followed the general guidelines for case study research by Yin [56] and the specific guidelines for case study research in empirical SE [14]. We used an inductive, bottom-up approach that began with analyzing decision makers' mental models when deciding the value of a feature. Data was collected through individual interviews with the key stakeholders involved in deciding upon features in our company case. GT's principles were applied in data analysis [19]. Note that we did not aim to generate a theory out of this single case study; however, this study sets the basis for future theory building by case study replications [57]. We did our best to define the case study context so to facilitate the identification of commonalities and differences between contexts. Appendix A details the steps we believe should be undertaken to replicate the study. Next, each phase of the research design is further developed.

### 3.1 Literature Review

We reviewed the literature in two phases. The study began with a *non-committal* literature review [19] to understand the topic and identify research gaps (presented in Sections 1 and 2). The second author conducted it, so to avoid the first author from being biased during data analysis. The non-committal literature review was based on the VBSE book [5] and a knowledgeable selection of high-quality papers on the topic. In a second phase, after analyzing the data, the two first authors conducted a deeper analysis of the literature to *integrate* the findings in the context of existing knowledge (as presented in Section 5). In this case, the literature was revised in a systematic way via a forward snowballing on [42], given that [42] had previously covered the related literature until 2009.

### 3.2 Case Selection and Context Description

Our case company is a large provider of telecommunications equipment that has a long tradition in software processes. In 2005, it moved to agile software development and created internal mechanisms to change their product-related decision-making processes so to include not only technical stakeholders, but also other key stakeholders representing areas such as marketing and business. Table 1 summarizes the case study context, based on the suggestions provided in [58] and [59].

The company develops several products. We focused on key stakeholders' value propositions (unit of analysis) within the features selection process for one of their key software intensive products. The product is used in different markets and distributed to end-users through intermediate companies that are the direct company customers. External releases are delivered approximately every six months. The study was conducted when the third release of the product was under negotiation.

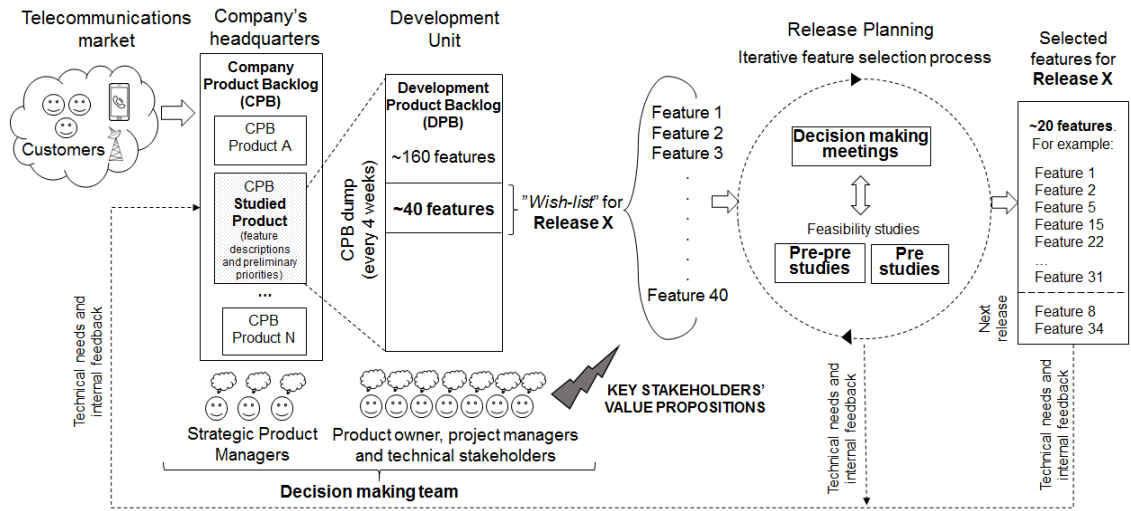

Fig. 1. Feature selection process at the company case.

The product is developed in two company locations: the company's headquarters, which is the main driver behind the set of features added to the Company's Product Backlog (see Figure 1), and a unit located in a different country that is responsible for the development (development unit herein). The development unit has around 300 people and follows an iterative agile process based on Scrum.

The feature selection process is also iterative (see Figure 1). Based on market/customers' requests, technical needs and internal feedback, high level features are defined by three strategic product managers (SPMs) that are located in the company's headquarters. Features are written in natural language and documented in a company level product backlog (CPB), which contains also features for other products. Only SPMs have access to the CPB. At this level, the information recorded per each feature is a brief description of the feature and a preliminary ranking based on the SPMs understanding of each feature's importance. Examples of features could be '*Real-time traffic optimization. Priority: high*' and '*Management system to full visibility of user performance. Priority: medium*'[1]. In addition, the development unit uses a development product backlog (DPB) to manage more detailed information regarding features. The DPB is a dump from the CPB, which contains details such as feature complexity, development cost or required knowledge to develop a feature. A dump from the CPB to the DPB is done every four weeks. In this way, updates in the CPB are transferred to the DPB as features are continuously discussed and priorities may change.

The product backlog for the product (represented in both, the CPB and the DPB) contains a '*wish-list*' with a window of two years, which roughly corresponds to 160 features. The feature wish list per release is limited to around 40 features. However, a release usually includes 20-25 features. We focused our study upon this later decision point. That is, when stakeholders have to decide upon the features that will be selected for the next release, based on the wish list of 40 features contemplated for that release.

Regarding the feature selection process, besides the three SPMs, additional stakeholders from the development unit are involved in decisions - the product owner, project managers and technical stakeholders. Feature selection meetings are periodically organized. In parallel, feasibility studies (pre-studies) are conducted to obtain more information about features.

Features are selected according to the value stakeholders believe these features will provide if added to the product. Within this context, decisions are carried out within a complex knowledge domain, of an uncertain nature [60] (i.e. knowledge is partially based on beliefs and cannot be assumed to be absolute with deterministic outcomes [61]). Power point presentations, pre-studies and customer studies are used during this process. Still, decisions are based on decision-makers' tacit knowledge [62] and knowledge is transferred via socialization [62] during feature selection meetings. Knowledge sharing via socialization is usually the case in agile software development [8]. The problem is that when knowledge is transferred on a tacit to tacit level, the end result is not concrete because value propositions remain hidden in decision-makers' minds [62]. A suitable solution to support decision-making under uncertainty is to build models that make explicit decision makers' mental models [63]. Following this way of thinking, the company's interest in the research falls in their desire to improve their decision making process by making it more systematic and transparent.

### 3.3 Subjects and Data Collection

Data was gathered through individual interviews with the key stakeholders responsible for deciding upon the target product's features (i.e. stakeholders participating in feature selection meetings). These stakeholders were part of the organizational structure and the only ones with executive power. They employ their own criteria and also represent the needs and suggestions from other stakeholders who do not directly participate in decision making meetings - internal and external stakeholders such as customers, development teams, maintenance teams, etc. A champion in the company (P10) identified the key stakeholders in our case, as shown in Table 2. They represent all key stakeholders involved in features selection for this product; therefore, we can claim that we have a complete picture of value in this specific case.

---
[1] Fictitious examples for illustrative purposes only.

TABLE 2
INTERVIEWEES' PROFILE

| P# | Role (SW and/or HW) | Experience in SW industry and Company A business (years) | Experience in current role | Interview length |
|---|---|---|---|---|
| P1 | Strategic product manager (SW) | 20 years (all of them in Company A) | 5 years | 66 minutes |
| P2 | Strategic product manager (SW) | 20 years (17 years in Company A) | 1.5 years | 70 minutes |
| P3 | Strategic product manager (SW and HW) | 25 years (all of them in Company A) | 2 years | 46 minutes |
| P4 | Product owner (SW) | 15 years (3 years in Company A) | 1 year | 84 minutes |
| P5 | Project manager (SW) | 20 years (1 year in Company A) | 10 months | 57 minutes |
| P6 | Project manager (SW) | 15 years (1 year in Company A) | 10 months | 28 minutes |
| P7 | Project manager (HW) | 19 years (9 months in Company A) | 9 months | 61 minutes |
| P8 | System verification manager (SW) | 27 years (20 years in Company A) | 2.5 years | 50 minutes |
| P9 | Continuous integration manager (SW) | 10 years (8 months in Company A) | 8 months | 49 minutes |
| P10 | Ecosystem manager (SW and HW) | 23 years (3 years in Company A) | 3 years | 53 minutes |

The interviewees were domain experts with a wide experience in the software industry, the business domain and conducting feature selection tasks (average of 19, 9 and 2 years respectively). Stakeholders in different roles participated in the interviews: SPMs (3), product owner (1) and different types of managers (6). Our focus was only on software features. Still, a hardware-related stakeholder, P7, was also interviewed because s/he is key to keep the hardware and software parts of the product well aligned when selecting features. In all cases, we made it clear upfront that the interview's focus was on software feature selection.

*Strategic product managers (P1-P3)*. SPMs are responsible for assigning resources to features. Their duties focus on strategic product decisions from business goals and product roadmaps to feature prioritization. SPMs are in constant contact with customers, with the business intelligence unit and with the strategic management group. Thus, it can be said that SPMs are the customer's voice for the product development. Among the three SPMs, P1 is the responsible for making the final decision based on discussions with the rest of key stakeholders during feature selection meetings. P1 and P2 focus on the software part of the product, while P3 is responsible of the overall solution.

*Product owner - PO (P4)*. The PO is the person in charge of prioritizing software features in the development unit. His/her role acts as a bridge between SPMs and development teams. In cooperation with the SPMs, the PO is responsible of maintaining the DPB.

*Project managers (P5-P7)*. The product has three project managers who are responsible to deliver the actual product out of their projects. Two project managers deal with software related projects (P5 and P6), whilst another project manager focuses on hardware related projects (P7). Project managers meet the SPMs and the PO once per week to provide feedback on how features' development is progressing as well as features that, from their point of view, should be part of the next release.

*System verification and continuous integration (CI) managers (P8-P9)*. P8 and P9 are close to the technical product development. P8 is responsible for the verification and validation teams and the lab equipment needed to conduct end-to-end product verification and validation. P9 is the CI manager and leads the agile development framework at the development unit.

*Ecosystem manager (P10)*. P10 manages partnerships that the unit establishes with other industries to create common interest ecosystems. S/he oversees all high-level decisions done in the development unit, including also decisions relating to the studied product.

Interviews were semi-structured to give room for participants to recall the value propositions they use in the context of feature selection. All interviews were planned to last for approximately one hour. We designed an interview script (provided in Appendix B) composed of three sections: 1) warm-up questions, including demographic and context setting questions (5 min); 2) value propositions elicitation questions, to gather knowledge about the set of value propositions that the interviewee uses when deciding upon a feature (45 min); and 3) wrap-up questions, to ascertain any missing relevant topic that the interviewee wanted to discuss (10 min). In addition to the interview's script, we used a drawing describing the feature selection process, as it appears in Figure 1, to make it clear that the interview would focus upon the decision point in which the 40 software features of the release wish-list had to be reduced to 20-25 features. The actual interview length is showed in Table 2. Although few interviews lasted less than one hour (e.g. P6), we did not find that the information gathered lacked quality. All interviews were face-to-face, voice recorded and transcribed. The first two authors conducted them between March/May 2015.

### 3.4 Data Analysis

The interview transcripts went through an iterative multi-step process of data analysis in a systematic manner using coding techniques from the Glaserian version of GT (open and selective coding) [64], [19]. The six phases described in thematic analysis by Cruzes and Dybå [65] were also considered during the analysis. However, we did not code based on large chunks of data (themes), but coded at a deeper level of detail[2]. We applied the constant comparison method (CCM) [19] and, therefore, the analysis was iterative. However, to achieve a legible description, it is described in a sequential manner.

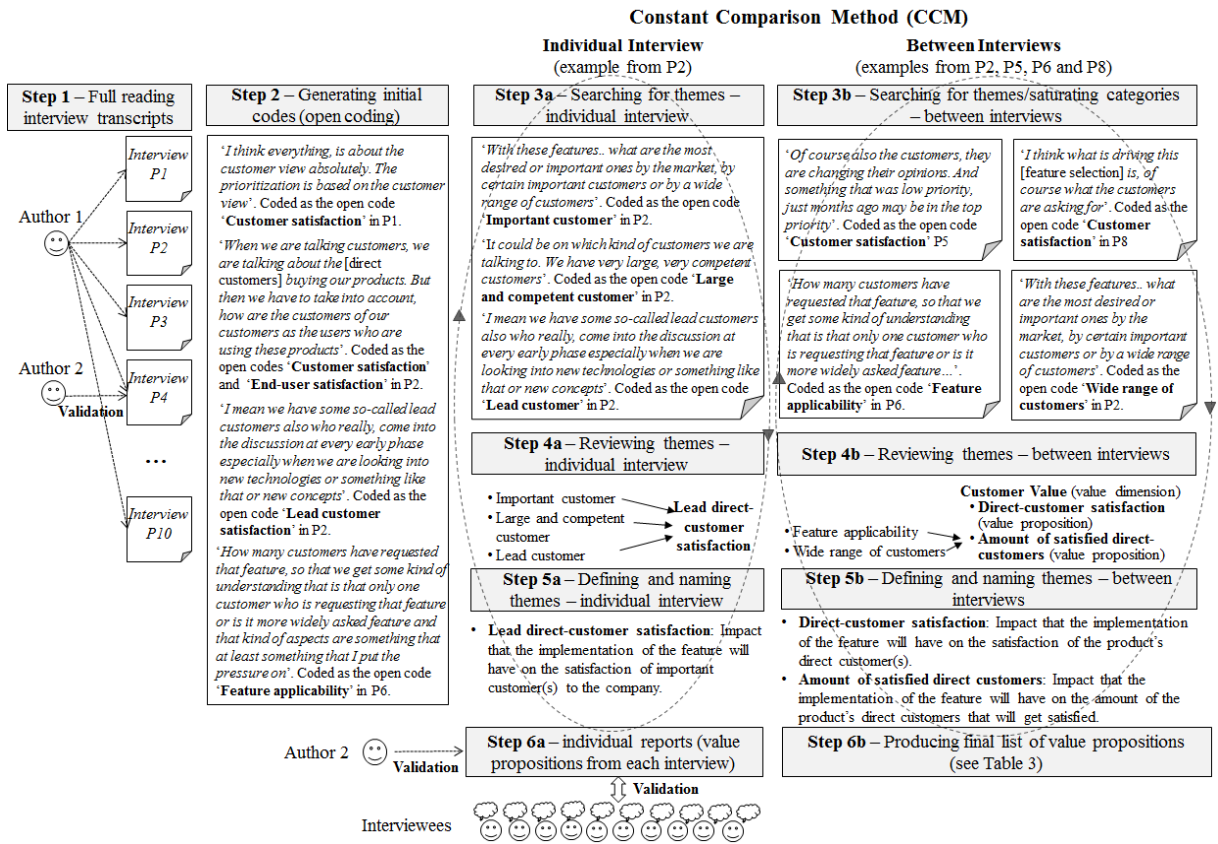

Fig. 2. Iterative data analysis adapted from Grounded Theory's coding techniques [19].

Figure 2 depicts the process using, as an illustrative example, one of the value dimensions identified in the study, 'customer value'. The first two authors of the paper – A1 and A2, participated in the analysis. First, A1 read each interview transcript to obtain an overview of the data (step 1, Figure 2). Then, she analyzed each transcript inductively. First, she used open coding (step 2, Figure 2) by going through the data, line by line, and attaching codes to relevant concepts (in this case value propositions). The CCM was used to identify patterns within individual interviews and selective coding was applied to cluster open codes around value propositions and value dimensions. Figure 2 shows some instances of the value proposition 'lead direct-customer satisfaction' as they were coded in P2's interview transcript. P2 referred to customers that were close to the company using terms such as 'important customers', 'large and competent customers' and 'lead customers'. We aggregated all under the value proposition 'lead direct-customer satisfaction'. By applying steps 3a, 4a, 5a and 6a, value propositions were identified from each interview and emailed back to each interviewee for validation.

Next, A1 applied the CCM further to identify patterns between interviews and saturate categories (step 3b). Themes were reviewed and modified when needed (step 4b) until achieving the final value propositions and dimensions (step 5b), as presented in Table 3 (step 6b).

In order to mitigate researcher bias in the analysis both A1 and A2 analyzed interview P4 separately (P4 was the first interview carried out). Before starting the coding process, A1 and A2 agreed upon what a value proposition represented to minimize researcher subjectivity during coding. Value propositions were assumed to be aspects that interviewees believed to be impacted upon if a feature was to be added into the release. To ensure validity, A1 and A2 went through interview P4 together and compared their respective individual coding. The goal was to check whether both researchers identified a similar list of value propositions. The level of agreement was high (75% - from the 16 value propositions identified from interview P4, 12 propositions were commonly identified by both researchers). The main differences were in the terminology used to code value propositions. It was agreed that, for all the remaining interviews, A1 would analyze them and prepare individual reports with the list of propositions and A2 would review the set of propositions identified in each interview prior emailing them to the interviewees. To keep a clear chain of evidence, the reports emailed to interviewees included the list of value propositions and definitions, and the quotations from the interviews that supported each proposition. The qualitative analysis tool NVivo[3] was used in the analysis.

---

[2] Urquhart [19] describes the relationship between GT and thematic analysis.

[3] http://www.qsrinternational.com/

# 4 RESULTS

This section presents the results by addressing the RQs. First, RQ1 is answered by analyzing the value propositions that were identified in the interviews. Then, we present a reconciliation of stakeholders' value propositions, by analyzing alignments and distinct views between them (RQ2).

## 4.1 Research Question 1: Value Propositions

We identified 36 value propositions, which were classified into six dimensions (grouping of value propositions that refer to a similar aspect):
1) Customer value
2) Market competitiveness
3) Economic value/profitability
4) Cost efficiency
5) Technology & architecture
6) Company strategy

The customer value dimension encompasses propositions deemed relevant from a customer's perspective. Market competitiveness comprises propositions relating to a product's competitiveness in the market, in terms of added value compared to similar products. The economic value/profitability dimension includes propositions relating to the impact that a feature would have to the company's wealth, mainly in monetary terms, in the long run. Cost efficiency propositions focus on the several costs that implementing a feature would entail. The technology & architecture dimension centers around technical aspects such as product architecture and existing capability to implement a feature. Finally, the company strategy dimension focuses on aspects that provide overall direction to the company in terms of organizational goals and strategies to achieve these goals.

Table 3 shows the list of value propositions, their description, and the corresponding number of stakeholders who mentioned them. Note that descriptions emerged directly from the interviews. As shown in Table 3, some value propositions (e.g. customer satisfaction, ROI) are rather general, likely applicable to many other companies, and in some cases, already identified in previous studies (see Table 6). They are also grounded on our empirical data; thus being genuinely used by decision makers in our case. Other propositions are more specific to our company case; for example, both customers and end-users are distinct entities, with different roles, and equally important to our company's business.

Note that not all value propositions were discussed in the same level of detail. The Totals column in Table 3 shows the number of interviewees that mentioned each proposition as well as the corresponding number of times (instances) that the proposition was identified in the transcripts. Propositions belonging to the customer value dimension were the ones mentioned the most, i.e. coded 143 times in NVivo. The dimensions mentioned the least were economic value/profitability and company's strategy. With regard to the latter, seven stakeholders made reference to topics related to this dimension and 39 instances were recorded in NVivo. As for economic value/profitability, it was mentioned by six interviewees, who stressed its importance because '*the company wants to make revenue in any case*' (P2, SPM). However, we only found 34 corresponding instances in NVivo.

Next, we use illustrative quotations from the interviews to provide supporting evidence for value propositions and dimensions. Illustrative quotations help keep a clear chain of evidence and support readers' understanding of the phenomenon [19], [66]. Note that to comply with our NDA, we made slight modifications on some quotations. Whenever this occurs, we use the following tag: [*modified text*].

### 4.1.1 Customer value

The Customer value dimension encompasses seven value propositions (F1) to (F7), all driven one way or another by the value a feature will provide to customers. Our analysis showed that interviewees considered four types of customers (F1-F4): company's direct customer(s), lead direct customer(s), product's end-users, and the market in general. Although one can argue some overlap between different customer types (e.g. company's direct customers and lead direct customers are part of the market in which the company operates), they were not merged so to stay genuine to what the interviewees mentioned. Direct customers, lead direct customers, company's business market and product's end-users have respectively 67, 24, 25 and 13 instances in NVivo. Next, we provide some examples of the evidence upon which our results on types of customers are grounded.

**Direct customer-related evidence (F1)**: P3 (SPM) described that '*we discuss with customers and we have customers' requirements […], customers tell us what they would like to have also in the future. So, I think the main thing is really what do we know about, what have the customer told us, how they will evolve* [the product] *for example. That's one very important input parameter*'. Similarly, P4 (PO) explained that '*the customer's view becomes the first. SPM* [who provides the customer's view] *has always the responsibility of ruling out something or bringing something as mandatory in*'. Features that either enable or disable a customer case are prioritized high: '*Let's say for example, this one customer wants to* [use the product]. *If they want to utilize our equipment, then we need to be able to comply with* [certain customer's characteristics]. *Otherwise, they will not be able to utilize our stuff. So, that kind of feature goes to the top*' P4 (PO).

**Lead direct customer-related evidence (F2)**: During the product's first release, features' assessment was quite straightforward as feature selection was based upon a single customer's needs. However, as the number of customers increased (current situation), the feedback of **lead customer(s)** became the one primarily followed: '*We typically select, let's say, ten most important customers that we have and we focus on those. We don't talk to all customers. We have* [x] *customers. We talk to maybe, 10 or 15, the biggest ones […]. Then, we believe that eventually everyone will use* [the feature]' (P1, SPM).

**End-users related evidence (F3)**: '*When we are talking customers, we are talking about* [direct customers] *buying our products. But then, we have to take into account how are the customers of our customers as the users who are using these products. So I think it's always a two-stage approach*' (P2, SPM).

TABLE 3
ANSWER TO RQ1: VALUE PROPOSITIONS (VALUE FACTORS)

| Id | Value Proposition | Description | Totals[3] |
|---|---|---|---|
| *Customer value* | | | **10 (143)** |
| F1 | Direct-customer satisfaction | Impact that the implementation of the feature will have upon the satisfaction of customer(s) (i.e. to what extent the feature is desired/needed by company's direct customers). | 10 (67) |
| F2 | Lead direct-customer satisfaction | Impact that the implementation of the feature will have upon the satisfaction of important customer(s) to the company (i.e. to what extent the feature is desired/needed by certain important customers). | 5 (24) |
| F3 | End-user satisfaction | Impact that the implementation of the feature will have upon the satisfaction of end-users (i.e. to what extent the feature is desired/needed by the product's end-users). | 4 (13) |
| F4 | Market satisfaction | Impact that the implementation of the feature will have upon the market satisfaction (i.e. to what extent the feature will be accepted by the market in general in which the company operates). | 7 (25) |
| F5 | Amount of satisfied direct-customers | Impact that the implementation of the feature will have upon the amount of company's direct customers that will get satisfied (i.e. amount of company's direct customers that desire/need the feature). If the feature is requested by a high number of customers, then the feature will be more valuable. | 4 (8) |
| F6 | Direct-customer value proposition | Impact that the implementation of the feature will have upon the customer's value proposition (i.e. how value is delivered to direct customers through the product). | 3 (3) |
| F7 | Overall direct-customer solution | Impact that the implementation of the feature will have upon the overall solution that the company offers to its customers (i.e. to what extent the feature fits well with other products that the direct customer uses from the company – customer's portfolio). | 2 (3) |
| *Market competitiveness* | | | **10 (111)** |
| F8 | Product competitiveness | Impact that the implementation of the feature will have upon the product's situation regarding to competitors (i.e. will the company get a competitive advantage if the feature is implemented? will the implementation of the feature support the company's technology leadership?). | 5 (21) |
| F9 | Product quality | Impact that the implementation of the feature will have upon the quality of the product (it may be also negative, for example, if the feature is very complex from a technical point of view). | 6 (25) |
| F10 | Product performance | Impact that the implementation of the feature will have upon the performance of the product (e.g. number of users handled by the product). | 4 (8) |
| F11 | Product simplicity | Impact that the implementation of the feature will have upon the simplicity of using the product from a customers' point of view (e.g. simplicity to maintain the product). | 2 (2) |
| F12 | Product ubiquity | Impact that the implementation of the feature will have upon the product's capability to support existing services in a transparent way. | 1 (3) |
| F13 | Time-to-market | Impact that the implementation of the feature will have upon the time-to-market of the release (i.e. time when the release is available for sale). It includes aspects such as the risk associated with delaying the release if the feature is included and the extent to which is possible to implement the feature on time so it will provide a competitive advantage. | 9 (46) |
| F14 | Overall release content | To what extent the implementation of the feature will contribute to provide a compact package to be released (e.g. value of the feature when considering also the other features that will be included in the release). | 4 (6) |
| *Economic value/profitability* | | | **6 (34)** |
| F15 | ROI | Feature's return on investment in monetary terms (i.e. worth brings by the feature if it is compared to its development cost). | 5 (16) |
| F16 | Sales volume | Impact that the implementation of the feature will have on the number of units of the product that will be sold (e.g. product's licenses). | 2 (10) |
| F17 | Customer base/ market share | Impact that the implementation of the feature will have on the market share (e.g. whether the implementation of the feature will help engage new customers). | 2 (3) |
| F18 | Product's price | Impact that the implementation of the feature will have upon the price of the product. | 1 (1) |
| F19 | Time-to-profit | Feature's time-to-profit (i.e. when the company will start making profit out of the feature). | 1 (1) |
| F20 | Opportunity cost | Impact that the implementation of the feature will have upon opportunity cost (i.e. economic benefits that could have been received by selecting an alternative feature). | 2 (3) |
| *Cost efficiency* | | | **10 (88)** |
| F21 | Overall development cost | Overall cost/effort of implementing the feature (including design, implementation, testing, verification and integration work, etc.). | 10 (55) |
| F22 | Testing cost | Cost/effort of testing the feature (low level testing). | 5 (10) |
| F23 | Integration and verification cost | Cost/effort of the complete integration and verification of the feature (end-to-end verification). | 3 (10) |
| F24 | Testing equipment/ environment cost | Impact that the implementation of the feature will have upon the need to invest in testing equipment to be able to test the feature. | 3 (9) |
| F25 | Subcontractor investment | Impact that implementing the feature will have upon making investments on third-party products or services (subcontractors). | 2 (2) |
| F26 | Reusing existing content | Impact that the implementation of the feature will have upon reusing existing content (i.e. to what extent the feature can be implemented by reusing existing content of the same product or other company's products). Reusing existing content will have a positive impact in decreasing costs. | 2 (2) |

[3] Number of stakeholders that mentioned the factor in the interview (number of instances recorded in NVivo for the factor).

TABLE 3 (CONTINUATION)
ANSWER TO RQ1: VALUE PROPOSITIONS (VALUE FACTORS)

| Id | Value Proposition | Description | Totals |
|---|---|---|---|
| *Technology and architecture* | | | **9 (82)** |
| F27 | Implementation complexity | Impact that the implementation of the feature will have upon the complexity of implementing the product (e.g. if there are some limitations or challenges in the complexity or technical dependencies). | 8 (25) |
| F28 | Product's architecture & internal quality | Impact that the implementation of the feature will have upon the architecture of the product (i.e. to what extent the implementation of the feature will enhance the internal quality of the product). | 4 (5) |
| F29 | Hardware impact | Impact that implementing the feature will have from a hardware point of view (e.g. whether implementing the feature will require changes on the hardware and that will increase the cost of the product). | 3 (14) |
| F30 | Technical relevance | To what extent the feature is relevant from a technical point of view (e.g. whether the feature is important from a technical perspective). | 3 (5) |
| F31 | Development capability | Impact that the implementation of the feature will have upon the capability of the organization for developing this feature and other features, considering both knowledge and resources (e.g. whether the timing of the feature from a resources/competences point of view is good or not). | 9 (33) |
| *Company strategy* | | | **7 (39)** |
| F32 | Company's market strategy | Impact that the implementation of the feature will have upon supporting the company's market strategy (i.e. to what extent the implementation of the feature complies with the current strategy of the company). | 4 (13) |
| F33 | Product's particular strategy | Impact that the implementation of the feature will have upon building the concept of the product that fits into the product's strategy (i.e. for example, considering how solutions like this product are being built and what kind of outlooks are in the market). | 3 (6) |
| F34 | Company's portfolio | Impact that the implementation of the feature will have upon the company's portfolio as the total offering of the company (i.e. to what extent the feature fits in/adds value to the company portfolio and it is well aligned with the features of other company's products). | 5 (14) |
| F35 | Business enabler | Impact that the implementation of the feature will have upon boosting other business, products or functionalities and enabling new value streams in the context of the product (e.g. boosting the sales of other company's products, maintaining existing businesses, etc.). | 3 (5) |
| F36 | Brand | Impact that the implementation of the feature will have upon the company as a brand (it includes aspects such as environmental impact, as the contribution that the company is making to sustainability by implementing the feature, and socio-economic impact, as the socio-economic contribution that the company is making by implementing the feature). | 1 (1) |

SPMs (P1 and P2) also stated that engineers should understand how to translate end-user needs into technical problems. For example, P1 (SPM) described that '*all these services that [the product] provides to the end user, those have to be supported […] that is maybe the most important part of the prioritizing*'.

**Market satisfaction-related evidence (F4)**: P1 (SPM) referred to the market as follows: '*so we go talk to customers about what they want and what we will provide, and we look at the market and so on. Then we collectively try to decide what to do next*'. An additional way to consider the market as a consumer relates to features that must comply with certain market regulations. '*There are certain mechanisms in different countries and certain regulations that we have to provide certain features which we cannot sell. They just have to be there. Sometimes countries or regions like the European Union just change their mind or come up with new regulations where they feel this is important to secure the security of users […]. We have to just follow these regulations because otherwise we cannot anymore sell the product in certain markets*' (P2, SPM).

Next, we provide grounded evidence relating to the three remaining value propositions under the Customer value dimension.

**Amount of satisfied direct customers-related evidence (F5)**: Customer value is not only about what features will satisfy customers, but also about how many customers will get satisfied by a feature. For example, P2 (SPM) said that they pay attention to whether a feature '*is something that we actually can do business of, even if it's a more customer-specific functionality or mechanism, so that it can be applied also to other customers*'. Similarly, P6 (project manager) mentioned that '*how many customers have requested that feature, so that we get some kind of understanding that it is only one customer who is requesting that feature or it is more widely asked feature. That kind of aspects are something that at least I put the pressure on*'.

**Direct customer value proposition-related evidence (F6)**: This value proposition represents the company's interpretation about the way in which value is delivered to and experienced by the customer. For example, P1 explained '*we prioritize whatever that customer needs, or what we think it is. I mean, we don't just follow what the customer says. […] sometimes, for example, you have these very basic things […]. But maybe they don't even think about it. But I know they need this feature*' (P1, SPM). In the same way, P8 (system verification manager) said, '*we have an idea of what we can provide to the customers, which they don't know yet but they will gain money on that, and we will gain money on that. I think it's naïve to think that you can go and ask the customers what they want because they don't know*'.

**Overall direct customer solution-related evidence (F7)**: This value proposition refers to the impact that the implementation of a feature will have on the overall solution that the company provides to its customers. P10 (ecosystem manager) explained it in the following way: '*It's not only [this product], but it's the whole thing what we are offering. […] you cannot only look that feature this and that feature that, you need to look at, OK, if I do these features in [this product], are they supported in [other company's products] as* *well?, because* [customers], *often when they build their* [solution], *they use* [example of different company's products that work together as an overall solution]. *So therefore, there is always interdependence between different products so they have to be part of the same kind of solution'*.

*4.1.2 Market competitiveness*

The dimension Market competitiveness includes seven value propositions (F8) to (F14) that refer to the ability of the company to successfully compete through the selection of features. As done for the previous dimension, here we also provide grounded evidence for each proposition.

**Product competitiveness-related evidence (F8)**: This proposition relates to the added value, compared to similar products already available in the market, that the implementation of a feature will bring. P3 (SPM) described that '*what we think the competition is doing in this area is a fairly important input as well. So we try to look a little bit on what competition is doing and try to see what can be done*'. Features are checked against competitors' products to assess whether the product has any gap compared to competitors. P2 (SPM) explained that the company has '*special people looking into the competition as well […], trying to understand what is the competition up to, what kind of products they are trying to launch, what is their strategy behind these products, what are the strengths and weaknesses of ours compared to theirs*'.

**Product quality-related evidence (F9)**: Before a feature is selected, '*things that could have certain impact on the initial quality of the product*' (P2, SPM) are analyzed. Features that either improve product quality or that are product enhancements to noticed problems are considered valuable. '*If we have some quality problem that could be helped with some new features* […] *or maybe we get a lot of faults or returns because of something that we could fix with a new software feature, then of course that would be something to... I: So then one factor is if this feature is going to improve the current product? R: Yes. Absolutely*' (P2, SPM). Note that although many interviewees saw quality as a general aspect, others were more specific, seeing quality as product performance (F10), product simplicity (F11) and product ubiquity (F12).

**Product performance-related evidence (F10)**: Four interviewees mentioned this proposition. P2 (SPM): '*sometimes we're getting some kind of field measurements also how the actual performance is against to what we have specified in the beginning.* […] *sometimes there are areas where we are, afterwards, realizing that there are problems and we have to come up with either an enhancement or an additional functionality, actually, to overcome these kind of problems*'. Similarly, P6 (project manager) stated that some discussions when selecting features revolved around '*if you could do this kind of requirement and change that algorithm such a way you could give a better performance*'.

**Product simplicity-based evidence (F11)**: Two interviewees mentioned that customer experience is improved if the product, from a customers' point of view, is kept simple. P1 (SPM): '*from* [a customer] *point of view, to be able to make it simpler for them, to deploy and maintain the product. Those types of features…, that is something that is of interest*'.

**Product ubiquity-related evidence (F12)**: This value proposition represents service transparency, which was considered a must so end-users see that services work smoothly. This proposition was mentioned by P1, who emphasized its importance: '*these services has to be ubiquitous today* […] *so that is maybe the most important part*'.

**Time-to-market-related evidence (F13)**: This value proposition was considered crucial to obtain competitive advantage. Indeed, it was mentioned by nine interviewees and collected 46 instances in NVivo. For example, P10 (ecosystem manager) explained that '*time to market, is obviously something that, especially product management should think and they certainly do, that OK, if we do this with these parameters, it is pointing us to come to the market at this time. Are we early compared to competition or need? There are two aspects: if we are early on competition and if that meets the client need, then it is good. If we are early on competition but client doesn't need it, it doesn't bring any value*'. In general, the right time in which a feature should be delivered is discussed when selecting features, P2 (SPM): '*do we believe that there is a business case to make money out of them* [features]*?, and, if the answer is yes, then okay we should look into what is the next suitable market when is the next release* [suitable for the feature]'.

**Overall release content-based evidence (F14)**: Four interviewees stressed the importance of assessing features considering a release's overall content, rather than to focus solely upon individual features. P6 (project manager) explained how it was a situation in which they '*decided that okay, we're not doing those features but we do instead something different, because if we want to keep this kind of pretty compact package* [the release], *to make that out into that time schedule, we are forced to postpone it*'.

*4.1.3 Economic value/profitability*

The Economic value/profitability dimension contains six value propositions (F15 to F20); it represents the profit that the company will make as a result of including a feature. A total of six interviewees mentioned it during the interviews, resulting in 34 instances in NVivo.

**ROI-based evidence (F15)**: When analyzing aspects related to the monetary worth (monetary value) of a feature, we found out that most interviewees pointed out return-on-investment (ROI) (i.e. overall profitability of a feature when considering the monetary worth that the feature will bring versus its development costs). For example, P4 (PO) explained that '*many kinds of technical things might end up in a situation where you think that this is actually not bringing too much extra value, if you compare to the cost*'. P2 (SPM) also mentioned: '*of course, I have to take into account what are the development costs versus the sales I can make out of that. How is the pricing? Do I have a business case here really?*'

Hereafter, we describe the value propositions that appeared concerning economic value. Note that development costs, which were widely discussed during the interviews, are added to Cost efficiency, in Section 4.1.4.

**Sales volume-related evidence (F16)**: Sales volume refers to the number of product units that can be sold if a feature is implemented. P1 (SPM): '*if we can increase the number of licenses of* [the product] *that we sell to each cus-* *tomer, that's of course also important'*. In addition, sales volume is also used to understand trends in the market. '*Next generation of the same product, of course, we can already see the trend, how the sales are developing, in what kind of characteristics of the product are important there in reality*' P2 (SPM). Sales volume is closely related to the next value proposition, Customer base/market share.

**Customer base/market share-related evidence (F17):** This value proposition refers to the ability of a feature to engage new customers. For example, P1 described the case of a software feature for which he was pushing to, even if the development unit argued that they could not implement it on time. P1 insisted that the feature should be implemented, even if other features would be sacrificed. When we asked the reasons why that feature was so important to him, he replied '*because it will enable us to target a lot more customers. So, it will sell more* [product's licenses] *basically. Interviewer: So, then, increasing customer base would be one of the points? P1: Absolutely*'.

**Evidence for Product price (F18), Time-to-profit (F19)** and **Opportunity cost (F20)**. These value propositions were mentioned by few interviewees and, therefore, they are not as saturated as previous value propositions (i.e. we did not find many instances in the data related to them). Still, they were included in this dimension because they were mentioned by P1 and P2 (two of the three SPMs). Related to the price of the product, P2 mentioned it as a variable affecting the economic worth brought by a feature. P2 (SPM) also made reference to Time-to-profit: '*Will I get a turnaround within a suitable time which we have standardized within the company? There has to be a turnaround, of course, but, when are we start making money with that feature?*' In addition, Opportunity cost, as the benefit that could have been received by taking an alternative feature, was mentioned by P1 (SPM) and P8 (system verification manager). P8: '*the product management thinks, yeah, we need to have that because that will blow, that will be a showcase. But we know it's going to be really complicated […]. It's gonna mean that we won't be able to do so many other things*'.

### 4.1.4 Cost efficiency

Costs were some of the most mentioned aspects during the interviews (all interviewees made reference to costs one way or another, and this aspect was coded 88 times in NVivo). Most interviewees focused on a feature's overall development cost (F21). However, some of the technical stakeholders referred to testing costs as impacting features' implementation (e.g. testing cost (F22), integration and verification cost (F23) and testing equipment/environment cost (F24)). Please note that interviewees referred to costs as representing economic cost, effort, and capabilities. For example, P10 (ecosystem manager) said '*how much do I need to invest? Resources, competencies, money, you name it*', P1 (SPM) put special emphasis on effort '*It's the effort. So actually money doesn't really matter*', and P5 (project manager) pointed to it in monetary terms '*Work estimate, it all goes back to dollar signs and how much effort it will take for my crew*'.

**Overall development cost-related evidence (F21):** Denotes the global cost/effort of implementing a feature, including design, implementation, testing, and whatever costs associated with a feature development. When asking P1 (SPM) '*what do you need to know from the* [development unit]*?*' S/he replied, '*I think, the most important thing is the cost. Interviewer: For each feature? R: Yes*'. Similarly, P4 (PO) explained that besides feedback on the technical solution, the second feedback that s/he provides to SPMs is the cost: '*Something might be just too costly, or it's more costly than you think comparing to the other content competing on getting to the release. So you might favor some two small features and you take some big out*'.

**Testing costs-related evidence (F22)**: This value proposition was emphasized by five interviewees (P4, P5, P7-P9). P5 (project manager) explained that '*a very big effort in this kind of projects is testing*'. An illustrative example of how testing costs impact the development of features was given by P9 (continuous integration manager): '*Are we capable of developing these features in our main track, so the active development track, or is there that sort of aspects in some features that requires for example their own software development track, which is, of course, then overhead for us in terms of the resources and test environments? […]. If you double or triple the amount of tracks, then we need more inhering the code from track to another. We need testing in all of those tracks […]. Therefore, it generates more costs*'.

**Integration and verification cost-related evidence (F23)**: P4 (PO) described that '*the final testing, the integration and verification, takes usually quite long time in calendar-wise […]. That's one of the biggest impacts*'. Similarly, when we asked P8 (system verification manager) about the kind of discussions that s/he has with SPMs, s/he explained that: '*in our team, we will do the verification on that feature […]. And, in that process, to give an estimation of how many hours do we need to spend on this, and whether we need to buy extra testing equipment etc. to be able to test that feature*'.

**Testing equipment cost-related evidence (F24)**: This value proposition was also mentioned by P7 and P9. For example, P9 (continuous integration leader) mentioned: '*There was a testing related discussion in a way that, what sort of test environments we need, how expensive those test environments are, test setup. Because different features requires probably…, the other feature requires double the price of the testing environment than the other one. So, what is our testing capacity and what it should be with this kind of feature?*'

**Subcontractor investment (F25) & Reusing existing content-related evidence (F26)**: These two value propositions had a low number of instances in the data (two for each). Regarding **Subcontractor investment (F25)**, some features are partially or totally implemented by subcontractors, which generates subcontractor costs. P2 (SPM) mentioned '*do we have to buy in third-party products to make this implementation happening?*' Two interviewees mentioned **Reusing existing content (F26)** as a strategy to decrease development cost and, also, as a proposition to be considered when selecting features. For example, P5 (project manager) explained that '*if there's something that I can reuse, I will reuse […]. Of course, I need to have also my sensors out there that somebody has almost done something similar so I* *will of course take, utilize all of that. Interviewer: is that something discussed when features are being selected? P5: Yeah I would say […]. For example,* [a concrete part of the product] *pretty much they took their previous product and just brought the same content to another environment'*.

*4.1.5 Technology and architecture*

This dimension encompasses technical aspects to be considered when selecting features. Five value propositions (F27 to F31) were identified related to this dimension.

**Implementation complexity-related evidence (F27)**: This value proposition was judged important to consider when selecting features due to its possible impact upon product's functionality, quality and time-to-market. SPMs described how they communicate with the development unit in this sense. For example, P2 mentioned that '*I need at least a high-level understanding of how they would like to implement and build the product, and how this impacts on certain functionality, if there are some limitations or some challenges in the complexity, which might delay the product launch, if there are things that could have some certain impact on the initial quality* (P2, SPM). P4 (PO) also explained how complexity is evaluated in the development unit: '*when we start analyzing the feature we need to understand what kind of impact it has for our sub-systems […]. We have to evaluate that where is the impact of this functionality, so, what areas are impacted. Maybe one of them, maybe all of them at some most complex features*'.

**Product's architecture & internal quality-related evidence (F28)**: Besides P4 (PO), who commented on the internal impact of the feature in the product's subsystems, P8 also mentioned that '*some features are pretty simple and some features are really complicated to put in there and get them working without demolishing the rest of the system*'. This is the main reason why sidetracks are sometimes needed, which increases the cost of the implementation (see Section 4.1.4 Cost efficiency). Features designed specially to improve internal quality were mentioned by P9 (continuous integration manager): '*For example, we have this kind of implementation in our architecture and it hasn't been required for three years. So, then, should we refine that? Should we probably make an own feature, which is kind of an improvement item? Not a commercial feature for the customer but the internal architecture change that enables us to make more cost-efficient R&D.*' However, P5 (project manager) pointed out that these kinds of discussions only briefly lead the selection of features: '*we may touch it briefly, but it's not the major issue there*'.

**Hardware impact-related evidence (F29)**: This value proposition represents the hardware side of the product, which can also be impacted by a feature's implementation. It was mentioned that hardware becomes sometimes a restriction for the product. '*If you're talking about features,* [P4] *is more or less caught in the* [hardware] *we have. He can't really, when it comes to the software it's hard to add things from this feature discussion on the hardware*' (P3, SPM). Similarly, P2 (SPM) mentioned that '*we need to take into consideration what can we do with the hardware in the software market*' and P5 (project manager) commented that '*for example, the customer is just asking something that your hardware platform is not supporting. So, you just cannot make whatever you want if it's just not sitting in that platform*'.

**Technical relevance-related evidence (F30)**: This value proposition was suggested because customers sometimes request features that are relevant in the context of other company's products but not in the context of this product. P4 (PO) explained that '*we have a feature here or wanted functionality, for example, from the customer perspective. But, after evaluating how it would need to be done, it might end up to be more like a not so important in the end product perspective*'.

**Development capability-related evidence (F31)**: This value proposition represents the development unit's capability for developing a feature, which is related to resources allocation (i.e. what kind of competences are needed for implementing the feature and whether those competences are available). Development capability captured the focus of many discussions during the interviews (nine interviewees mentioned it and it added to 33 instances in NVivo). For example, P5 (project manager) explained that '*the input that comes from* [P4] *and myself is that typically, there's one team that becomes the bottleneck […]. We say that hey, now if we pick and choose this feature then, we are 100 per cent utilizing* [Team A]. *So that's it, we can't have anything more. So, we are mainly looking into what can be done*'. P4 (PO) corroborated this perspective: '*sometimes one of the teams or the sub-areas may be the bottleneck so that that might be also the reason why some feature even if it's higher in the priority, it's not anymore possible*'.

*4.1.6 Company strategy*

The last dimension relates to the strategy that the company follows for the development of its products, which does not include only this specific product, but all the products in the company's portfolio. This dimension includes five value propositions, F32 to F36.

**Company's market strategy-related evidence (F32)**: This value proposition refers to the overall strategy that the company follows to develop its products. For example, P2 (SPM) explained that '*of course, every company has certain strategy where we want to go. Features, products, which markets to address and things like that. I have to take that, of course, into account as well […]. I cannot come up with a wild idea and say this will make millions of money, if it does not comply with our company strategy*'. Furthermore, P2 elaborated that '*if it does not comply with our strategy […], then, there could be alternatives which we can discuss with the customer, but in worst case we have to say no, we don't do that*'.

**Product's particular strategy-related evidence (F33)**: This value proposition was mentioned by some interviewees relating to scenarios when some features are announced as part of the product's marketing strategy and get priority. '*We were talking a lot about this* [specific feature]. *So that is, we have already gone out with the market message that this is the best* [product] *in the world […], it can do* [specific feature] *[…]. We demonstrated this* [specific feature], *but it was not really available to customers. Now, because of that, that is the most important thing to do*' P1 (SPM).

**Company's portfolio-related evidence (F34)**: This proposition refers to the total offering of the company. Usually, customers do not buy exclusively this product but a wider solution (see F7). Thus, features need to be well aligned across the entire portfolio, so that the overall solution provided to customers works as a whole and differentiates this Company from its competitors. P10 (ecosystem manager) explained that '*it doesn't make sense that you do something for this particular product only, but it doesn't fit to the total offering that we have. And that doesn't exclude features*'. P1 (SPM) made reference to the importance of catching up with existing features in other company's products. '*This [product] is the new addition to our product portfolio and it has a separate software stack compared to our other products. So, that makes this process maybe a little bit different than for other products because we have this gap, in terms of what features are supported*'. P3 (SPM) also explained that '*in [this product] case, we know what we are doing in [other products] and we try to sort of, should we have the same set of features or capabilities to some extent what we do in [other products]. That's another source of input*'.

TABLE 4
ANSWER TO RQ2: OVERLAPPING, SPECIFIC TO A STAKEHOLDER GROUP AND CONFLICTING VALUE PROPOSITIONS

| Overlapping value propositions | Value propositions specific to a stakeholder group | Differing value propositions |
|---|---|---|
| Value propositions mentioned by each stakeholder group and almost all stakeholders:<br>− Direct-customer satisfaction (F1)<br>− Time-to-Market (F13)<br>− Overall development costs (F21)<br>− Implementation complexity (F27)<br>− Development capability (F31) | Value propositions that provide details on aspects that are relevant from the stakeholder group's expertise. In particular:<br>− *SPMs (P1-P3)*: value propositions related to Customer value, Market competitiveness, Economic value/profitability, and Company strategy.<br>− *Project managers (P5-P7)*: value propositions related to Cost efficiency. In particular, strategies to decrease costs.<br>− *Technical managers (P8-P9)*: value propositions related to Cost efficiency and Technology & architecture. In particular, testing costs and impact in product architecture. | No evidence of conflicting views on what value means in the context of feature selection for the studied product (i.e. value propositions). However:<br>− The view on some value propositions differed between some stakeholders. In particular, Testing related costs (F22, F23 and F24) and Implementation complexity (F27).<br>− There were also some divergence with regard to what is seen as a feature's acceptable quality level (F9). |

**Business enabler-related evidence (F35)**: This proposition relates to a scenario where a feature to be decided upon is likely not to provide direct revenue but is likely to help boost other businesses or maintain current business. For example, P2 (SPM) explained that '*we have to earn money, and to make money. But sometimes, of course, it could be also that you say okay we are launching this feature even if we are not making, but we are now, we are boosting perhaps another product with that. That could also be a case [...]. We can boost the sales of another product or another functionality*'. Similarly, P10 (ecosystem manager) explained that '*when you look at it once again, in the bigger picture, you need to look at also that yeah, this feature itself doesn't bring us much value, but enables us to do something else, where we can get new value stream*'.

**Brand-related evidence (F36)**: This value proposition represents the Company as a brand and was mentioned by P10. Particularly, s/he stated that '*in my opinion we always need to think about the brand as well. One is environmental impact and one is social or socio-economic impact*'.

Note that the value propositions detailed herein were not all orthogonal. Our focus was to identify the propositions deemed suitable and meaningful to our key stakeholders, even if this led to factors that are correlated (e.g. ROI, sales volume and overall development cost).

### 4.2 Research Question 2: Stakeholders Analysis

Once value propositions were identified, our next step was to investigate how different stakeholder groups understood value to answer RQ2: *What are the overlapping and distinct value propositions (value factors) between the different stakeholder groups involved in value-based feature selection?* There were five stakeholder groups participating in feature selection for the target product: Strategic Product Managers (SPM), Product Owner (PO), Project Managers (PM), Technical Managers (TM) (system verification and continuous integration managers) and Ecosystem Manager (EM) (see Section 3.3). Table 4 summarizes the answer to RQ2, distinguishing between stakeholder groups' overlapping and group-specific value propositions (distinct value propositions); the third column focuses on differing views between stakeholder groups with regard to some propositions.

**Value propositions common to all stakeholder groups (and almost all stakeholders)**, are *Direct-customer satisfaction* (F1, 10/10) (Customer value dimension), *Time-to-market* (F13, 9/10) (Market competitiveness dimension), *Overall development cost* (F21, 10/10) (Cost efficiency dimension), *Implementation complexity* (F27, 8/10) and *Development capability* (F31, 9/10) (Technology & architecture dimension). The two dimensions that were not common to all stakeholder groups were Economic value/profitability, and Company strategy; these were mostly driven by SPMs.

**Specific value propositions** were identified on dimensions that are closer to the stakeholder group's expertise. They were identified for three different stakeholder groups: propositions related to Customer value, Market competitiveness, Economic value/profitability and Company strategy for SPMs; propositions related to Cost efficiency for PMs; and propositions related to Cost efficiency, and Technology & architecture for TMs.

For **conflicting views between stakeholder groups**, results show that, in general, interviewees had a well-aligned understanding of value when selecting features. For example, there were no dimensions mentioned by a single stakeholder group and completely ignored by other groups. In addition, we did not find evidence that could suggest contradictory views for most value propositions; particularly, aspects such as what the customer wants (Customer value propositions) or the strategy that the product should follow (Company strategy propositions) were well-aligned. However, we identified different views in aspects related to Implementation complexity (F27), testing related costs (F22, F23 and F24), Product quality (F9) and Time-to-market (F13). TMs stressed testing related costs – aspect that did not emerge in the interviews with SPMs. Likewise, TMs mentioned the need to evaluate in more detail a feature complexity before its selection. Regarding Product quality and Time-to-market, we noticed that whilst TMs aimed at high quality features all the time, SPMs referred to a level of quality that was *good enough* to deliver the product on time.

Next, we further detail the way in which these findings emerged from the analysis. Table 5 shows the value dimensions and propositions as pointed out by stakeholder groups and also by specific stakeholders. As Table 5 - Total column shows, most stakeholders made reference to the six value dimensions in one way or another. Except from the Economic value/profitability and Company strategy dimensions, where their corresponding value propositions were referenced by six and seven stakeholders respectively, all the remaining dimensions were considered by all stakeholders, i.e., all stakeholders mentioned value propositions that would fall within those dimensions. However, the number of propositions provided to specific value dimensions differed depending on the stakeholder group. That is, specific stakeholder groups stressed specific value dimensions and detailed their value propositions further.

**SPMs:** SPMs (particularly P1 and P2) were the stakeholders who contributed with the highest number of value propositions (22 and 27 respectively). Such results may suggest that SPMs have a wider overall view of the decision criteria used for feature selection, when compared to other stakeholders, who tend to narrow their views according to their areas of expertise. For example, all stakeholders mentioned Direct-customer satisfaction (F1). However, whilst most interviewees cited only a few value propositions belonging to Customer value, both P1 and P2 heavily discussed customer-related topics. A similar situation happens with the value propositions belonging to Market competitiveness; Product competitiveness (F8) was mainly discussed by SPMs and aspects related to Product quality (F9, F10, F11 and F12) were more often mentioned by SPMs and the PO (P4). Regarding Economic value/profitability, except for ROI, which was mentioned by five interviewees, mainly SPMs pointed out to this dimension. The same happens with Company strategy, where propositions were mainly referred to by SPMs, together with the EM.

**PMs and TMs**: PMs and TMs provided a detailed view on aspects in the value dimensions Cost efficiency and Technology & architecture. For example, with regard to Cost efficiency, whilst SPMs only discussed overall development cost, without pointing out to specific costs, PMs and TMs were more concrete and mentioned the relevance of testing costs for features' implementation; in particular, P8, P9 and the PO (P4). On the other hand, reusing existing content (F26), as a strategy for reducing cost, was only mentioned by PMs (P5 and P6), which is reasonable as they try to reduce costs to get the maximum benefit out of the budget assigned to their projects. Similarly, regarding the value dimension Technology & architecture, SPMs discussed Implementation complexity (F27) in general, whilst stakeholders from the development unit (P4, P6, P8 and P9) were more concrete and pointed out the impact upon the product's architecture and its subsystems (F28).

**PO:** When analyzing the value propositions mentioned by the PO (P4), we noticed a link between SPMs and stakeholders in the development unit. The PO has a good understanding of the value propositions that are important for SPMs (those related to Customer value and Market competitiveness) and also a reasonable visibility of technical aspects related to Costs and Technology & architecture (aspects that were more frequently mentioned by technical stakeholders). Except for the Company strategy dimension, for which P4 did not mention any value proposition, value propositions mentioned by P4 are fairly distributed among the different dimensions. For example, although P4 did not mention all propositions belonging to Customer value and Market competitiveness, s/he presented a fair understanding of aspects that are important in these dimensions (3/7 and 4/7 propositions respectively). Similarly, s/he mentioned cost and technical aspects in more detail than SPMs (3/6 and 5/6 respectively).

**EM:** Finally, when analyzing the view of the EM, we noticed that his/her view is closer to the SPMs' view. S/he emphasized propositions belonging to the dimensions Customer value and Company strategy (4/7 and 4/5 respectively), and provided only an overall view on propositions related to Cost efficiency and Technology & architecture (1/6 and 2/6).

Overall, it seems that the understanding of value between different stakeholder groups is not divergent but complementary, so different stakeholder groups provide different perspectives in the decision process via their different stakeholders' profiles.

## 5 DISCUSSION

This section integrates our results into the existing body of knowledge on the topic. Our goal is to analyze to what extent the value propositions identified in our study have been also considered in related work. We also discuss implications to research and practice and threats to validity.

### 5.1 Comparison to Value Propositions Identified in Related Empirical Studies

One of the chapters of the VBSE book (2006) [5] presents a study, claimed to be the first that analyzes criteria for selecting requirements in the context of VBSE [13]. Existing empirical research in this area is relatively scarce, though. We found eight studies that report *empirical* findings relating to criteria[4] used when assessing the value of a feature/requirement[5].

---

[4] Related studies use different terms when referring to value propositions such as decision-making criteria, value factors, value aspects, etc.

[5] Related studies do not make a clear difference when using the terms requirement and feature, despite emphasizing requirements in titles and abstracts (e.g. [12], [45]).

TABLE 5
STAKEHOLDERS' VALUE PROPOSITIONS RECONCILIATION

| Value Factor | Strategic Product Managers (SPMs) | | | Product Owner | Project Managers | | | Syst. Ver. manager | CI manager | Ecosystem manager | TOTAL |
|---|---|---|---|---|---|---|---|---|---|---|---|
| | P1 (SW) | P2 (SW) | P3 (SW/HW) | P4 (SW) | P5 (SW) | P6 (SW) | P7 (HW) | P8 (SW) | P9 (SW) | P10 (SW/HW) | |
| ***Customer value*** | x | x | x | x | x | x | x | x | x | x | **10** |
| F1 Direct-customer satisfaction | 1 | 1 | 1 | 1 | 1 | 1 | 1 | 1 | 1 | 1 | 10 |
| F2 Lead direct-customer satisfaction | 1 | 1 | | 1 | 1 | 1 | | | | | 5 |
| F3 End-user satisfaction | 1 | 1 | | | | 1 | | | | 1 | 4 |
| F4 Market satisfaction | 1 | 1 | 1 | 1 | | 1 | 1 | | | 1 | 7 |
| F5 Amount of satisfied direct-customers | 1 | 1 | | | | 1 | | 1 | | | 4 |
| F6 Direct-customer value proposition | 1 | 1 | | | | | | 1 | | | 3 |
| F7 Overall direct-customer solution | | 1 | | | | | | | | 1 | 2 |
| *Customer value - Total* | 6 | 7 | 2 | 3 | 2 | 5 | 2 | 3 | 1 | 4 | |
| ***Market competitiveness*** | x | x | x | x | x | x | x | x | x | x | **10** |
| F8 Product competitiveness | 1 | 1 | 1 | | | | 1 | | | 1 | 5 |
| F9 Product quality | 1 | 1 | 1 | 1 | | | 1 | | 1 | | 6 |
| F10 Product performance | 1 | 1 | | 1 | | 1 | | | | | 4 |
| F11 Product simplicity | 1 | | | | | | | 1 | | | 2 |
| F12 Product ubiquity | 1 | | | | | | | | | | 1 |
| F13 Time-to-market | 1 | 1 | 1 | 1 | 1 | 1 | 1 | 1 | | 1 | 9 |
| F14 Overall release content | | 1 | 1 | 1 | | 1 | | | | | 4 |
| *Market competitiveness - Total* | 6 | 5 | 4 | 4 | 1 | 3 | 3 | 2 | 1 | 2 | |
| ***Economic value/profitability*** | x | x | - | x | x | - | - | x | - | x | **6** |
| F15 ROI | -1 | 1 | | 1 | 1 | | | 1 | | 1 | 5 |
| F16 Sales volume | 1 | 1 | | | | | | | | | 2 |
| F17 Customer base/market share | 1 | 1 | | | | | | | | | 2 |
| F18 Product's price | | 1 | | | | | | | | | 1 |
| F19 Time-to-profit | | 1 | | | | | | | | | 1 |
| F20 Opportunity cost | 1 | | | | | | | 1 | | | 2 |
| *Economic value/profitability - Total* | 3 | 5 | 0 | 1 | 1 | 0 | 0 | 2 | 0 | 1 | |
| ***Cost efficiency*** | x | x | x | x | x | x | x | x | x | x | **10** |
| F21 Overall development cost | 1 | 1 | 1 | 1 | 1 | 1 | 1 | 1 | 1 | 1 | 10 |
| F22 Testing cost | | | | 1 | 1 | | 1 | 1 | 1 | | 5 |
| F23 Integration and verification cost | | | | 1 | | | | 1 | 1 | | 3 |
| F24 Testing equipment/environment cost | | | | | | | 1 | 1 | 1 | | 3 |
| F25 Subcontractor investment | | 1 | | | | 1 | | | | | 2 |
| F26 Reusing existing content | | | | | 1 | 1 | | | | | 2 |
| *Cost efficiency - Total* | 1 | 2 | 1 | 3 | 3 | 3 | 3 | 4 | 4 | 1 | |
| ***Technology and architecture*** | x | x | x | x | x | x | x | x | x | x | **10** |
| F27 Implementation complexity | 1 | 1 | 1 | 1 | | | 1 | 1 | 1 | 1 | 8 |
| F28 Product's architecture & internal qu. | | | | 1 | -1 | 1 | | 1 | 1 | | 4 |
| F29 Hardware impact | | 1 | 1 | 1 | | | | | | | 3 |
| F30 Technical relevance | 1 | 1 | | 1 | | | | | | | 3 |
| F31 Development capability | 1 | 1 | | 1 | 1 | 1 | 1 | 1 | 1 | 1 | 9 |
| *Technology and architecture - Total* | 3 | 4 | 1 | 5 | 1 | 3 | 2 | 3 | 3 | 2 | |
| ***Company strategy*** | x | x | x | - | x | x | - | x | - | x | **7** |
| F32 Company's market strategy | 1 | 1 | | | | | | 1 | | 1 | 4 |
| F33 Product's particular strategy | 1 | 1 | 1 | | | | | | | | 3 |
| F34 Company's portfolio | 1 | 1 | 1 | | | 1 | | | | 1 | 5 |
| F35 Business enabler | | 1 | | | 1 | | | | | 1 | 3 |
| F36 Brand | | | | | | | | | | 1 | 1 |
| *Company strategy - Total* | 3 | 4 | 2 | 0 | 1 | 1 | 0 | 1 | 0 | 4 | |
| ***TOTAL*** | 22 | 27 | 10 | 16 | 9 | 15 | 10 | 15 | 9 | 14 | |

Table 6 shows the correspondence between value propositions from these related studies and our study. Note that we kept original names for the propositions identified in related studies and used their descriptions in the papers to make a semantic correspondence to our propositions. For example, '*Requirement's issuer*' and '*stakeholder priority of requirement*' were associated to our value dimension Customer value because they were connected to customers and markets in the original studies. Propositions that did not show a clear correspondence to our value propositions were included as 'Other factors', at the end of Table 6.

A first series of six survey studies was published between 2005 and 2009 [13], [43], [12], [44], [45], [42]. In 2015, Latha and Suganthi [46] replicated the survey with companies in India. The survey instrument used in these studies included thirteen propositions that are considered important when deciding upon the value of a software requirement. Survey respondents were asked to rate the propositions' relative importance in present and future requirements decisions. As presented in Table 6, criteria slightly changed along the studies. We used bold in Table 6 to highlight value propositions that surveys' respondents rated as relevant. Further, propositions that were not part of the thirteen predefined criteria but were suggested by respondents are marked with '*'. The main strength of such studies is that they cover a wide range of companies and domains. However, their findings are limited due to the lack of details on value propositions, as findings are mainly based on close-ended questions, and answers are constrained by predefined choices.

Besides these surveys, a recent study by Alahyari et al. [10] analyzed how 14 agile software development organizations interpret value, using interviews as data collection method. The study did not specify the types of decisions being made, which in itself leads to several problems: i) if the type of decision is not documented (e.g. feature selection), key stakeholders that play a relevant role in that decision cannot be identified, which hinders one from obtaining a complete view of value. In this study, two stakeholders participated per organization - product owner and process responsible. Further, ii) we cannot be sure that the value propositions identified in this study are, indeed, applicable to feature selection. However, despite such problems, we also compared Alahyari et al.'s results with ours because product owners are responsible for feature selection in Agile. Therefore, we hypothesize that value propositions elicited from them are applicable to our case. Moreover, it seems to be the only empirical study so far that has also investigated the value concept using interviews.

Overall, from the 36 value propositions that emerged from our study, 17 were mentioned in related studies and 19 are unique to our study (highlighted by grey shading in Table 6). This does not mean that these 19 value propositions have not been mentioned in any other literature; however, they are not included in the existing lists of empirically-based value propositions that are referenced herein. In addition, eight value propositions were mentioned in related studies but did not appear in our interviews (last row in Table 6). When having a closer look at the comparison we can see that:

- **On the generalization of value propositions**: Previous surveys suggest that differences between companies regarding criteria applied for selecting requirements/features may be not so large [13], [43]. '*It is quite clear that the companies have very similar opinions regarding what is important when deciding whether or not to include a specific requirement in the next project or release. This makes the results even more interesting than if the companies had differing opinions, because it points to the possibility of a pattern, or common trend in views, across the software development industry*' [43]. To a certain extent, our study supports this assertion since all criteria considered as important in previous surveys were also mentioned in our interviews. Moreover, although only customer satisfaction was common to all studies, value propositions such as *market satisfaction, product competitiveness, product quality, time-to-market, ROI, development capability* and *product strategy* were considered important at least in three previous studies. However, as it is described next, we identified many other propositions that did not appear in previous studies, from which some seem to be context-dependent.

- **On new value propositions brought by our study/value propositions missing in our study:** There are several value propositions that were not mentioned in previous studies and are, therefore, an addition of our study; noticeably in the value dimensions Customer value, Economic value/profitability and Cost efficiency.
  - If we focus on the Customer value dimension, propositions such as *Amount of satisfied customers*, *Customer value proposition* and *Overall customer solution* were not mentioned in previous studies. Further, *end-users*, key for our company case, were only briefly mentioned in [10].
  - *ROI* was emphasized by most previous studies. However, we identified more concrete aspects in this area, such as *Sales volume* – also mentioned by interviewees in [12] and [45], *Customer base/market share*, *Product's price*, *Time-to-profit* and *Opportunity cost*. Moreover, our interviewees mentioned very concrete sources of cost, compared to related studies (F22-F26).

  In addition, previous studies included some propositions that did not emerge in our interviews (last row Table 6). From the series of surveys, only *Support for education and training* was found relevant in [46]. Alahyari et al. [10] also elicited four value propositions that did not appear in our study: 1) *Processes, ways of working, tools*, and 2) *To keep positive attitude, professionalism*, which are process-oriented value propositions, and 3) *Innovation, knowledge for organization*, and 4) *Knowledge on feature value for customers*, which refer to the knowledge acquired by an organization. This aspect, which is aligned with *Support for education and training*, could perhaps be important when selecting features.

  Overall, some value propositions may be context-dependent (e.g. *Overall customer solution* (F7), *Product ubiquity* (F12), *Support for education/training*) and some may not have appeared in related studies due to their shortcomings. Barney et al. [12], [45] pointed out that software product value is dependent on the context in which the software product exists. Our study seems to also support this claim. Further investigation from other software companies will provide the means to aggregate findings, and identify possible patterns across the software development industry.

- **On the relative importance of some value propositions:** The surveys suggest that business and management criteria are more important than technical concerns. Business and management aspects were widely discussed in our interviews (e.g. F1, F3, F13, F15, F31). However, technical aspects were extensively discussed as well (e.g. F27, 25 instances in NVivo), including new aspects such as technical relevance (F30) and hardware impact (F29). Our findings suggest that our company case keeps a balance between business, management and technical value propositions.

TABLE 6
STAKEHOLDERS' VALUE PROPOSITIONS INTEGRATION TO RELATED EMPIRICAL STUDIES

| Value Factors | Wohlin and Aurum, 2005a [13]<br># Companies: 2<br># Subjects: 13<br># Countries: - | Wohlin and Aurum, 2005b [43]<br># Companies: 9<br># Subjects: 33<br># Countries: - | Barney et al., 2006 [12]<br># Companies: 1<br># Subjects: -<br># Countries: 1 | Hu et al., 2006 [44]<br># Companies: 6<br># Subjects: 72<br># Countries: 3 | Barney et al., 2008 [45]<br># Companies: 3<br># Subjects: 26<br># Countries: 2 | Barney et al., 2009 [42]<br># Companies: 11<br># Subjects: 107<br># Countries: 3 | Latha and Suganthi, 2015 [46]<br># Companies: -<br># Subjects: 26<br>#Countries: 1 | Alahyari et al. 2017 [10]<br># Companies: 14<br># Subjects: 23<br># Countries: 1 |
|---|---|---|---|---|---|---|---|---|
| **Customer value**<br>F1 Direct-customer satisfaction<br>F2 Lead d-customer satisfact.<br>F3 End-user satisfaction<br>F4 Market satisfaction<br>F5 # of satisfied d-customers<br>F6 D-customer value proposit.<br>F7 Overall d-customer solution | *Customer value*<br>***Req.'s issuer***<br>***Stakeholder priority of req.*** | ***Req.'s issuer***<br>***Stakeholder priority of req.*** | ***Req.'s issuer***<br>***Stakeholder priority of req.*** | ***Customer satisf.***<br>*Req.'s issuer* | ***Req.'s issuer***<br>***Stakeholder priority of req.*** | ***Customer satisf.***<br>*Req.'s issuer* | ***Req.'s issuer***<br>***Stakeholder priority of req.*** | *Customer relat.*<br><br>*End-user performance, usability* |
| **Market competitiveness**<br>F8 Product competitiveness<br>F9 Product quality<br>F10 Product performance<br>F11 Product simplicity<br>F12 Product ubiquity | * Market technology trends<br>***Competitors*** | ***Competitors*** | * Competitive advantage<br>***Competitors*** | ***Competitors***<br>***SW features*** | ***Competitors*** | ***Competitors***<br>***SW features*** | ***Competitors*** | ***Competitiveness***<br>***Perceived qual./ actual quality, Functionality***<br>***End-user performance, Reliability*** |
| F13 Time-to-market<br>F14 Overall release content | ***Delivery date/ calendar time*** | ***Delivery date/ calendar time*** | *Delivery date/ calendar time* | ***Calendar time*** | ***Delivery date/ calendar time*** | *Calendar time* | ***Delivery date/ calendar time*** | ***Delivery process w.r.t. time*** |
| **Economic value/profitability**<br>F15 ROI<br>F16 Sales volume<br>F17 Customer base/market share<br>F18 Product's price<br>F19 Time-to-profit<br>F20 Opportunity cost | ***Development cost-benefit*** | ***Development cost-benefit*** | ***Development cost-benefit*** | | ***Development cost-benefit*** | | *Development cost benefit* | ***Revenue, business value*** |
| **Cost efficiency**<br>F21 Overall development cost | * Product cost | | | ***Development cost***<br>* Total ownership costs | | *Development cost* | | ***Cost*** |
| F22 Testing cost<br>F23 Integration/verification cost<br>F24 Testing eqpt. /env. cost<br>F25 Subcontractor investment<br>F26 Reusing existing content | | | | | | | | |
| **Technology and architecture**<br>F27 Implementation complexity | *Complexity*<br>*Req. dependency.* | *Complexity*<br>*Req. dependency* | *Complexity*<br>*Req. dependency* | *Complexity*<br>*Req. dependency* | *Complexity*<br>*Req. dependency* | *Complexity*<br>*Req. dependency* | *Complexity*<br>*Req. dependency* | |
| F28 Product architecture & internal quality | *System impact*<br>*Evolution*<br>*Maintenance* | *System impact*<br>*Evolution*<br>*Maintenance* | *System impact*<br>*Evolution*<br>*Maintenance*<br>*Operating architecture* | *Evolution* | *System impact*<br>*Evolution*<br>*Maintenance* | *Evolution* | *System impact*<br>*Evolution*<br>*Maintenance* | ***Non-functional req., Hedonic, Maintainability*** |
| F29 Hardware impact<br>F30 Technical relevance | | | | | | | | |
| F31 Development capability | *Resources/ competences* | *Resources/ competences* | *Resources/ competences* | *Resources* | *Resources/ competences* | *Resources* | *Resources and competences* | |
| **Company strategy**<br>F32 Company's market strategy | * Strategic importance/alignment | | | ***Business strat.*** | | ***Business strat.*** | | |
| F33 Product's particular strategy<br>F34 Company's portfolio | * Function is promised/sold | | ***Function is promised/sold*** | | ***Function is promised/sold*** | | ***Function is promised/sold*** | |
| F35 Business enabler<br>F36 Brand | | | *New business | | | | | |
| **Other factors** (factors that did not emerge in our study) | *Req.'s volatility*<br>*Support for education/training* | *Req.'s volatility*<br>*Support for education/training* | *Volatility*<br>*Support/education/training*<br>*Business model*<br>*Adherence to design parameters* | *Req. volatility*<br>*Extra cost*<br>*After-sale support*<br>*Industry character* | *Volatility*<br>*Support/education/training* | *Req. volatility*<br>*Extra cost*<br>*After-sale support* | *Req. volatility*<br>***Support for education/training*** | *WoW and Tools To keep positive attitude…*<br>*Innovation, knowledge of organization*<br>*Knowledge of feature value for customer* |

**Proposition**: propositions that the study found relevant for selecting requirements/features (at least for one of the studied cases) have been highlighted in bold.
(*): propositions that were not part of the criteria predefined by researchers but were suggested by the survey's respondents are marked with *.

In addition, a difference between our results and previous surveys relates to *market competiveness*, which value propositions were heavily discussed in our interviews, but not stressed in the surveys. They appear in [10] though. Thus, it seems that taking into consideration competitors is becoming increasingly important. A similar situation happens regarding *Company strategy*. This dimension was not mentioned in [10], perhaps, because of the characteristics of the stakeholders included in that study (product owners and process responsible). Regarding the survey studies, besides, *Business strategy* [44] [42], and *Product's strategy*, which is partially considered in the criteria *Functions is promised/sold* in [12], [45] [46], aspects related to strategic issues were not stressed.

- **On value propositions and agile software development:** A recent systematic literature review on agile requirements [67] found that the greatest challenge in managing requirements/features in Agile is that only focusing on business value may become problematic, as there might be other factors to consider. Our results show that, in practice, although propositions belonging to the customer value dimension have a high impact in the decision-making of companies using Agile (like our company case), there are many other propositions that affect decisions as well.
- **On the nature of the value propositions:** Some value propositions clearly provide value and should be maximized (e.g. *Customer satisfaction*, *ROI*). However, there are other aspects that are also considered to decide upon the value of a feature that could be considered as 'anti-values', from the point of view that they are factors to be minimized (e.g. costs).

Finally, although the SVM [4] is not included in Table 6 because it is based on a mapping study of the literature and its empirical evaluation is limited to a static example (without further use in real cases), we compared our results with its value aspects to analyze to what extent its value propositions overlapped with ours. Our value dimensions, *Customer value*, *Market competitiveness*, *Economic value/profitability*, *Cost efficiency* and *Technology & architecture* are embedded in the SVM. However, the value dimension *Company strategy* is almost overlooked in the SVM. Similarly, the SVM's innovation and learning perspective emerged in our study from the point of view of product competitiveness, but 'learning' was not really emphasized during the interviews. Related to specific value propositions, some value propositions identified in our study are not present in the SVM (e.g. *product ubiquity*, *opportunity cost*, *subcontractor investment*, *hardware impact*).

### 5.2 Discussion on Stakeholder Analysis

The Theory W suggests the importance that satisfying all stakeholders has on the success of software projects [68]. Stakeholder value proposition elicitation and reconciliation is also an integral element that provides foundations for VBSE [3]. Overall, the literature stresses on different views of value, often incompatible, among the many stakeholders involved in software development [69] (e.g. developers would like to have stable requirements, but users want to be able to change product features when they see that a different feature would better satisfy their needs). A mismatch between decision criteria used by software developers and business stakeholders has been considered a major problem in the software industry [42], [45]. Wohlin and Aurum [13], [43] found also that although value propositions seem to be consistent among software companies, quite different opinions are given by individuals.

However, within the context of our specific case there is no such mismatch. We found some differences between stakeholders' views, particularly in terms of product quality and time-to-market. However, rather than divergent views between stakeholders, we identified complementary views that support a complete understanding of value. One reason for this might be that our company has already developed a culture of value in decision-making. Although the product is a new addition to the company portfolio, many of the stakeholders involved in deciding upon product's features have been working in the company for a long time, which might have created in them a company shared understanding of what is important when selecting features. Boehm [3] argues that conflicts between stakeholders are frequent in new situations but compatible value propositions can be achievable in situations of long-term stakeholder mutual understanding and trust.

The literature also shows sources of mismatch in quality, which impact upon product's time-to-market and costs. For example, Shaw [70] argues that features in software only need to be good enough to meet users' needs because users are commonly able to handle a certain level of unintended behavior, but bug fixing comes at a cost. We found a similar situation in our case. Our interviews revealed some differing views on these aspects, which suggests that they need particular consideration when reconciling stakeholders' value propositions.

### 5.3 Implications for Research and Practice

The knowledge on value provided in this study (i.e. six value dimensions, 36 value propositions and stakeholder analysis) aims to help researchers and practitioners learn about value-base feature selection based on rich data rather than on simplifications [66]. Our research elicited value propositions directly from key stakeholders' tacit knowledge through a context-based approach that allows for rich descriptions on value propositions and the rationale behind them. The current work raises several **research areas** that demand further consideration.

- Our results show agreement between our value propositions and value propositions identified in existing research (e.g. *customer satisfaction*, *product competitiveness*, *time-to-market*, *ROI*, *development capability* and *product strategy*), but also reveal several new value propositions (e.g. *amount of satisfied customers, customer base/market share, testing costs, time-to-profit*). Moreover, a significant amount of value propositions seems to be context dependent. Thus, value for feature selection seems to be diverse and both dependent on context and not. Our results set the basis for further theory building. We call for additional studies similar to ours to better understand possible patterns along value propositions and the fundamentals of those that are context dependent. Appendix A provides indications to facilitate the integration of new cases. It is essential that researchers clearly set-up the study's context, so to be able to aggregate results. Further, as much as possible, all key stakeholders involved in the specific phenomenon under investigation (e.g. feature selection) must participate in the research to ensure a complete view of value, which is not limited to only short-term aspects.

- Whilst some of our value propositions are well-saturated (i.e. we found many instances in the data that legitimize them), others have less instances in the data and further exploration of those is needed (e.g. *time-to-profit*, *pricing*, *opportunity cost* and *brand*).
- We suggest researchers to explore value propositions through research methods that allow for deep understanding such as the research method applied herein. The series of survey studies revealed the significant challenges that applying survey research in an emerging area of knowledge, such as value, brings. For example, these studies illustrated that is difficult to formulate an exhaustive set of criteria, particularly if criteria should be reasonably independent (e.g. practitioners had difficulties to understand the criteria as they are not fully orthogonal) [13][43][45]. Moreover, researchers had difficulties to draw conclusions as respondents tended to see relevance in all the criteria suggested by researchers [13][43][45]. Therefore, a bottom-up approach that can capture the details of the phenomenon seems to be more appropriate. Once the area is investigated further, and a more standardized terminology and detailing of value has been achieved, research methods such as surveys can be helpful in understanding trends.
- Our results are particularly important for researchers working on aspects related to software/software-intensive product management (e.g. release planning, feature management, and techniques and approaches for selecting features). The value propositions identified in this study can be used as an input for existing release planning methods. It is important that feature selection methods support value propositions identified from key stakeholders making decisions relating to the selection of those features; value propositions that were widely mentioned by our interviewees, and seem to be consistent among related studies are of particular interest.
- We noticed that, when describing the criteria used to judge the value of a feature, in some cases the stakeholders referred to qualities or non-functional requirements (NFRs) like complexity, simplicity and customer satisfaction. We preferred to document all value propositions without categorizing them further to stay genuine to what the interviewees mentioned, as there are some that are not NFRs (e.g. product competitiveness, sales volume). However, our results may be also of interest for the research community working on NFRs in aspects such as tradeoffs in NFRs, eliciting NFRs and prioritizing NFRs.
- Our results show that value propositions are not orthogonal, at least for our case. Understanding the relationship between them and whether all value propositions involved in feature selection have the same weight are fruitful areas for future research as well.

From a **practitioner's perspective** the list of value propositions identified in this study can be used as a comprehensive starting set to be adjusted to support feature selection in companies with similar characteristics to the company case described herein. We suggest an adjustment because our results showed that a significant amount of value propositions are context-dependent and, therefore, their suitability/revisiting will need to be considered in those companies' specific contexts. In addition, we believe that the details provided in the value propositions can help companies willing to employ value-based thinking to their decision-making processes. With regard to value dimensions, they are general but we believe they also provide a wider perspective that can be useful to companies that want to think in terms of value when selecting features. For example, one should use not only short-term aspects, such as development costs (dimension 4, Cost efficiency) or technical impacts (dimension 5, Technology & architecture), but also long-term aspects, such as Customer value (dimension 1), Market competitiveness (dimension 2), Economic profitability (dimension 3) and Company strategy (dimension 6). Note that companies need to ensure that decision making teams include all key stakeholders so that they represent as complete as possible picture of value. Moreover, they need to pay attention to possible sources of conflict between stakeholders. In our specific case, aspects related to feature's acceptable quality level. Regarding the use of value propositions during feature selection meetings, our results show that not all value propositions are common to all key stakeholders, i.e., different stakeholder groups focus upon different value propositions, and, within our results, they complement one another.

### 5.4 Validity Procedures and Validity Evaluation

This section discusses the study's validity threats in terms of construct validity, internal validity, external validity and reliability [14], [56]. Table 7 summarizes mitigating strategies for each of the validity threats.

*Construct validity:* The selection of the company case and instruments for data collection may influence the study's construct validity. Our company case already employed both short- and long-term aspects when selecting features; thus, it was aligned with the meaning of value thinking as per VBSE [3]. To mitigate misconceptions between interviewees and researchers (e.g. constructs discussed in the interview questions being interpreted in different ways), we organized four preparatory meetings with our champion (P10) and P4 (PO). During such meetings, they described the feature selection process, as drawn in Figure 1, and provided feedback on the terminology used in the interview. The analysis of individual interviews was emailed to participants to receive their feedback. P10, P4 and P9 provided feedback, which was considered in the analysis.

AUTHOR ET AL.: TITLE 21TABLE 7
PROVISIONS FOR SECURING TRUSTWORTHINESS OF THE STUDY

| Criteria | Description | Threats | Mitigation strategies |
|---|---|---|---|
| Construct validity | To what extent operational measures represent the concepts being studied according to the research questions. | – Relevance of the case to address the research questions.<br>– Rigor in data collection (e.g. misconceptions such as questions of the interview being interpreted in a different way by researchers and participants). | – Value-based decision making in place.<br>– Four walkthrough meetings with company representatives to set-up the study and prepare the interview script (~ 5 hours).<br>– Interviews recorded and transcribed.<br>– Analysis of individual interviews emailed to the interviewees for feedback. |
| Internal validity | The extent to which other confounding aspects/factors may influence the results that are identified. | – Interviewee's inaccurate view on value propositions. | – Selection of participants (all key stakeholders who participate in feature selection meetings).<br>– Very experienced participants. |
| External validity | To what extent the findings of the case study are of interest to other people outside the investigated case. | – Appropriateness of the company case.<br>– Representativeness of the company case. | – Company thinking in terms of value, all key stakeholders participated in the research.<br>– Detailed description of the study's context based on [68] [69].<br>– Steps to analyze/integrate additional cases in Appendix A.<br>– Detailed comparison to related studies. |
| Reliability | To what extent the study can be replicated obtaining the same results. | – Measurement bias (e.g. reliability of the measurement instrument/raw data).<br>– Researcher bias (e.g. preconceptions of researchers in data collection/analysis and inappropriate use of analysis methods).<br>– Participant bias (e.g. subject's subjectivity, willingness to provide reliable data). | – Interview script/questions prepared in accordance to the RQs.<br>– Interviews recorded and transcribed.<br>– Researcher triangulation in both data collection and analysis.<br>– Well-established coding techniques and tool support for data analysis (NVivo).<br>– CCM helped saturate value propositions and dimensions.<br>– Participants own motivation to be interviewed.<br>– Individual interviews, NDA's and guarantee for anonymity. |

A challenge when integrating our results into the existing body of knowledge was the different terminology used among papers and the lack of details in describing some propositions. Our analysis is based on the descriptions in the papers so to make a semantic correspondence between them and our propositions. However, our interpretation may have impacted the results in cases in which the original text was confusing.

*Internal validity:* We see as the main internal validity threat the selection of participants, as this may lead to the elicitation of value propositions that are not important when selecting features. Our company champion (P10) selected interviewees who represented all key stakeholders involved in selecting features for this product. All interviewees were very experienced in software development/management and had also a fair experience on value-based feature selection, which mitigates interviewee's inadequate view on value due to a lack of expertise. One could argue that the completeness of the results could still be threatened by the interviewees' ability to express their knowledge in words. However, all the interviewees received a detailed report with their corresponding value propositions, so to obtain further feedback. Further, by interviewing all key stakeholders the chances of missing relevant propositions became negligent.

*External validity:* To ensure that our findings are not only relevant for our company case but also allow others to understand how our results map to other contexts, care has been taken to describe the context of the study. Further, a detailed comparison to related studies is presented in Section 5.1, which increases the usefulness of the results. Appendix A details the steps other researchers would need to follow to analyze and integrate additional cases. Note, however, that the knowledge elicitation and data analysis techniques employed herein are equally applicable outside a VBSE context, thus showing a wider generalizability.

*Reliability:* We followed systematic procedures to guarantee the reliability of the evidence and minimize biased views. Based on the RQs, an interview script was prepared in advance. Although interviews were semi-structured, the same interview script was used in each interview. All interviews were recorded and transcribed. Researcher triangulation was used in both data collection and analysis, as well as well-established coding techniques (see Sections 3.3 and 3.4). The CCM helped us saturate categories, which strengths their legitimacy (see Table 3, Totals column). Our findings are based on how our interviewees perceive value (and their subjective opinions). However, given that feature selection is extensively based on tacit knowledge, we believe that the key stakeholders deciding upon software features for the studied product are the best source of knowledge to investigate value in our case. Interviewees were informed about their rights at the beginning of each interview; confidentiality was handled through NDAs and anonymity of individuals' responses was guaranteed.

## 6 CONCLUSIONS

This paper presents the results of a case study on value propositions employed for selecting software features, as well as how different stakeholder groups perceive value in the context of feature selection. 36 value propositions were identified, and classified into six dimensions: *Customer value*, *Market competitiveness*, *Economic value/profitability*, *Cost efficiency*, *Technology & architecture*, and *Company strategy*. Our analysis on how distinct stakeholder groups understand value shows that, in our company case, value perspectives from different groups are not divergent but complementary, representing different decision perspectives. Still, some sources of conflict were found in aspects related to product quality (i.e. what level of quality is enough).

The paper makes a number of important contributions. First, it sheds light into the important concept of value and value propositions used for selecting software features. Product value is represented by the set of features selected for its releases, hence the importance of understanding value in the context of feature selection. In practice, value is determined by a set of propositions that are not usually explicitly stated, but employed subjectively by decision-makers. One of our key findings is that, although the customer perspective has a crucial impact when selecting features, a wide spectrum of different propositions is considered. Further, value propositions such as *Direct-customer satisfaction*, *Time-to-market*, *Overall development cost*, *Implementation complexity* and *Development capability* seem to be unanimously agreed as critical factors when deciding upon software features.

Second, when integrating our results into the existing body of knowledge relating to our research, the results of our study are consistent with some of the results from previous research and suggest that it may be feasible to identify value patterns across different software industries. However some value propositions appeared also to be quite context dependent. Similar studies to ours will provide the means to aggregate data and to investigate whether patterns can be identified, as well as to extent the set of value propositions used for feature selection within similar contexts to ours.

Third, in our study, we applied coding techniques from the Glaserian version of GT, into a case study research, and gained significant insights that generated new value propositions, which have not been empirically identified in previous research. To our knowledge, this is the first study applying GT principles on VBSE that has been conducted in the SE discipline.

As described in Section 5.3, the findings of this study raises several research areas that need further research. We plan to continue our work in this line, aiming to theory building, and encourage other researchers to do so.

## ACKNOWLEDGMENT

We would like to thank the company for its commitment as part of our collaboration. This research has been carried out within the FiDiPro VALUE project number 40150/14, which is funded by Tekes (the Finnish Funding Agency for Technology and Innovation).

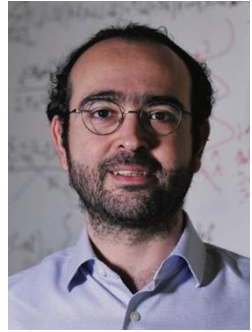

**Burak Turhan,** PhD (Bogazici), is an Associate Professor in Cyber Security & Systems at Monash University. His research focuses on empirical software engineering, software analytics, quality assurance and testing, human factors, and (agile) development processes. Dr. Turhan has published over 100 articles in international journals and conferences, received several best paper awards, and secured research funding exceeding 2 million euros. He has served on the program committees of over 30 academic conferences, on the editorial or review boards of several journals including IEEE Transactions on Software Engineering, Empirical Software Engineering and Journal of Systems and Software, and as (co-)chair for PROMISE'13, ESEM'17, and PROFES'17. He is currently a steering committee member for PROMISE and ESEM, and a member of ACM, ACM SIGSOFT, IEEE and IEEE Computer Society. For more information please visit: https://turhanb.net

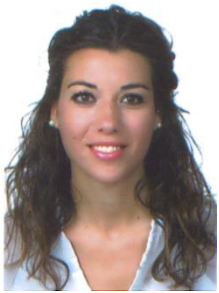

**Pilar Rodríguez.** PhD 2013 Computer Science; MSc. 2008 Computer Science; BSc 2006 Computer Science; Post-doctoral researcher at the Empirical Software Engineering in Software, Systems and Services (M3S) research unit of the University of Oulu (Finland). Her research centers on empirical software engineering with a special focus on value-based decision-making, agile and lean software development, software quality and human factors in software engineering. Dr. Rodríguez's research has been published in premier software engineering international journals and conferences. She is a member of the Review Board of EMSE (2015, 2016, 2017), and has served as reviewer in leading academic forums in Software Engineering (e.g. TSE, EMSE, IST, IEEE Software and ESEM).

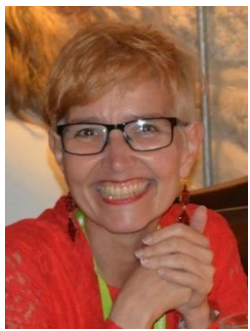

**Emilia Mendes.** PhD 1999 Computer Science; MSc. 1992 Computer Science; BSc 1986 Computer Science; Full Professor in Computer Science at the Blekinge Institute of Technology (Sweden), and also a Tekes-funded Finnish Distinguished Professor at the University of Oulu (Finland). Her areas of research interest are mainly within the context of empirical software engineering, value-based software engineering, and the use of machine learning techniques to contexts such as healthcare, and sustainability. She has published widely and over 200 refereed papers, plus two books as solo author – both in the area of cost estimation. She is on the editorial board of TSE, and has been former editorial board member in the journals SQJ and EMSE.